\documentclass[12pt]{article}

\usepackage[affil-it]{authblk}
\usepackage[numbers]{natbib}
\usepackage{graphicx}
\usepackage{amsmath}
\usepackage{amssymb}
\usepackage{theorem}
\usepackage{dcolumn}% Align table columns on decimal point
\usepackage{bm}% bold math
\usepackage{color}
\usepackage{soul}
\numberwithin{equation}{section}

{\theorembodyfont{\rmfamily}

}

\newcommand{\A}{{\boldsymbol{A}}}

\renewcommand*{\Xi}{{\boldsymbol{\xi}}}

\newcommand{\rbar}{{\bar{r}}}

\title{Model reduction for networks of coupled oscillators}

\author{Georg A. Gottwald}
\affil{School of Mathematics and Statistics, The University of Sydney, Sydney 2006 NSW, Australia
\thanks{georg.gottwald@sydney.edu.au}}

\date{\today}

\begin{document}

\maketitle

\begin{abstract}  
We present a collective coordinate approach to describe coupled phase oscillators. We apply the method to study synchronisation in a Kuramoto model. In our approach an $N$-dimensional Kuramoto model is reduced to an $n$-dimensional ordinary differential equation with $n\ll N$, constituting an immense reduction in complexity. The onset of both local and global synchronisation is reproduced to good numerical accuracy, and we are able to describe both soft and hard transitions. By introducing $2$ collective coordinates the approach is able to describe the interaction of two partially synchronised clusters in the case of bimodally distributed native frequencies. Furthermore, our approach allows us to accurately describe finite size scalings of the critical coupling strength. We corroborate our analytical results by comparing with numerical simulations of the Kuramoto model with all-to-all coupling networks for several distributions of the native frequencies.
%We present a collective coordinate approach to synchronisation which allows to study the onset of synchronisation in a Kuramoto model for phase oscillators. The $N$-dimensional Kuramoto model is reduced to a one-dimensional ordinary differential equation, constituting an immense reduction in complexity. Our approach allows to study the onset of synchronisation for finite number of oscillators as well as for infinite networks. We corroborate our analytical results by comparing with numerical simulations of the Kuramoto model with all-to-all coupling networks and Erd\H{o}s-R\'enyi graphs for several distributions of the native frequencies..
\end{abstract}

%\keywords{synchronisation; Kuramoto model; coupled oscillators; complex networks; collective coordinates}

\maketitle
 
{\bf Pacs numbers: 05.45.Xt, 89.75.-k, 89.75.Fb}\\ 
% 89.75.-k Complex systems 
% 05.45.Xt synchronisation; coupled oscillators 
% 89.75.Fb Structures  and organization of complex systems 
% PACS, the Physics and Astronomy Classification Scheme.

\begin{quotation}
Despite their inherent complexity large networks of interacting dynamical entities often exhibit coordinated ordered behaviour such as mutual synchronisation. The macroscopic behaviour of complex networks arises as a complicated interplay between the dynamics of each microscopic node and the overall topological properties of the network. It is a formidable challenge to reduce the dynamics of large networks to a small number of active degrees of freedom that is capable of capturing these complex dynamical phenomena. This work is a contribution towards this goal.
\end{quotation}
%%%%%%%%%%%%%%%%%%%%%%%%%%%%%%%%%%%%%%%
\section{Introduction}
\label{sec-intro}
The collective behaviour of interacting oscillators in complex networks is ubiquitous in nature and has occupied scientists from as disparate areas as biology, engineering, mathematics, physics and sociology for many years now \cite{Kuramoto,PikovskyEtAl,AcebronEtAl05,OsipovEtAl,ArenasEtAl08}. These systems often exhibit collective synchronisation whereby some or all oscillatory agents assume the same phase. Synchronisation behaviour is strongly dependent, amongst other factors, on the nature of the distribution of the native frequencies. In the case where all oscillators are connected with each other and where their native frequencies are unimodally distributed, for example, the onset of synchronisation as a function of the coupling strength is a soft transition, where the order parameter increases smoothly from zero as in a second-order phase transition. On the other hand, in the case of uniformly distributed frequencies, the onset of synchronisation is a hard transition, where at the critical coupling strength the order parameter has a non-zero value as in first-order phase transitions, with possible hysteresis \cite{Kuramoto,Pazo05,GomezGardenesEtAl07,LeyvaEtAl13}. Capturing all these different dynamic behaviours is a challenging task.\\

The collective behaviour of coupled oscillators such as synchronisation behaviour suggests that the dynamics of complex systems may (at least in certain cases) be described by a low dimensional dynamical system. To find these dimension-reduced descriptions is a formidable challenge with some remarkable results in recent years \cite{OttAntonson08,PikovskyRosenblum08,MarvelEtAl09,MartensEtAl09,PikovskyRosenblum11}. In this work we propose a new approach to describe coupled phase oscillators and their non-trivial dynamics. Our approach is not restricted to a thermodynamic limit of infinite many oscillators and allows for the study of finite size effects \cite{Pazo05,HildebrandtEtAl07,HongEtAl07,Tang11}, apparent in any real world networks.\\

The particular approach proposed in this work seeks to find an approximate parametrisation of the synchronisation manifold by means of appropriately chosen {\em{collective coordinates}} \cite{GottwaldKramer04,MenonGottwald05,MenonGottwald07,MenonGottwald09,CoxGottwald06}. The underlying premise is that the actual solution of the dynamical system assumes a specific functional form the parameters of which are coined collective coordinates. The temporal evolution of the actual solution is then described by the temporal evolution of those parameters, constituting an immense reduction in dimensionality. The functional form of the actual solution and the associated collective coordinates have to be specified upon inspection of numerical simulations of the underlying system. For the Kuramoto model we will establish that the phases are linearly correlated with the native frequencies and we define the collective coordinate to be the parameter relating the two. The method deals directly with the dynamical system rather than its associated macroscopic (infinite-dimensional) description for the distribution or moments thereof \cite{OttAntonson08,PikovskyRosenblum08,MarvelEtAl09,MartensEtAl09,PikovskyRosenblum11}. It is non-perturbative in the sense that the solution is not written as an expansion in some small parameter. The paper is organized as follows. In Section~\ref{sec-model} we introduce the Kuramoto model which constitutes a paradigm for studying coupled phase oscillators. Our approach to achieve effective model reduction of the dynamics is introduced in Section~\ref{sec-coll}. In Section~\ref{sec-numerics} the method is applied to the Kuramoto model with all-to-all coupling with three different distributions for the native frequencies and we compare the results of direct numerical simulations of the full Kuramoto model with those of the proposed $1$-(or $2$-)dimensional reduced model. We consider here a uniform native frequency distribution where a hard onset of synchronisation is experienced, a unimodal normal frequency distribution where a soft onset of synchronisation is experienced, and thirdly a bimodal frequency distribution where global synchronisation is preceded by partial synchronisation of weakly coupled synchronised communities. We conclude with a summary and discussion in Section~\ref{sec-summary}.

% finite size effects: Pazo05, Tang11, HildebrandtEtAl07, HongEtAl07
% low dimensional reductions:  \citet{OttAntonson08,PikovskyRosenblum08,MarvelEtAl09,MartensEtAl09,PikovskyRosenblum11}
% Caveat: when is this possible (see Martens EtAl paper for formulations and chaos)

%%%%%%%%%%%%%%%%%%%%%%%%%%%%%%%%%%%%%%%

\section{Kuramoto model}
\label{sec-model}
Weakly coupled limit cycle oscillators can be described in terms of their phases as an autonomous dynamical system. A widely used model which governs the dynamics of the phases $\varphi_i$ of $N$ oscillators with native frequencies $\omega_i$ is the celebrated Kuramoto model \cite{Kuramoto,Strogatz00,AcebronEtAl05}
\begin{align} 
\label{e.kuramoto}
{\dot{\varphi}}_i = \omega_i + \frac{K}{N} \sum_{j=1}^N a_{ij}\sin(\varphi_j-\varphi_i)\, .
\end{align}
The adjacency matrix $\A=\{a_{ij}\}$ determines the topology of the network and describes which oscillators are connected. We restrict our analysis to unweighted, undirected networks for which the adjacency matrix $\A=\{a_{ij}\}$ is symmetric with $a_{ij}=a_{ji}=1$ if there is an edge between oscillators $i$ and $j$, and $a_{ij}=0$ otherwise. The degree of a node $d_i$, i.e. the number of edges emanating from node $i$, is then given by $d_i = \sum_j a_{ij}$. 

For interacting oscillators, generically there exists a critical coupling strength $K_c$ such that for sufficiently large coupling strength $K>K_c$ the oscillators synchronise in the sense that they become locked to their mutual mean frequency and their phases become localized about their mean phase \cite{Kuramoto,OsipovEtAl,Strogatz00}. This type of synchronous behaviour known as global synchronisation occurs if the dynamics settles on a globally attracting manifold \cite{Crawford94}. The level of synchronisation is often characterised by the order parameter \cite{Kuramoto}
\begin{align}
\label{e.r}
r(t)=\frac{1}{N}|\sum_{j=1}^Ne^{i\varphi_j(t)}|\; ,
\end{align} 
with $0\le r \le 1$. In practice, the asymptotic limit of this order parameter 
\begin{align}
\label{e.rbar}
{\bar{r}} = \lim_{T\to\infty}\frac{1}{T}\int_{T_0}^{T_0+T} r(t)\, dt\; ,
\end{align}
is estimated whereby $T_0$ is chosen sufficiently large to eliminate transient behaviour of the oscillators. 

In the case of full synchronisation with $\varphi_i(t)=\varphi_j(t)$ for all pairs $i,j$ and for all times $t$ we obtain $\rbar=r=1$. In the case where all oscillators behave independently with random initial conditions $\rbar = \mathcal{O}(1/\sqrt{N})$ indicates incoherent phase dynamics; values inbetween indicate partial coherence. 

%%%%%%%%%%%%%%%%%%%%%%%%%%%%%%%%%%%%%%%
\section{Collective coordinate approach}
\label{sec-coll}
We will employ a non-perturbative approach to study synchronisation. Our approach is borrowed from the theory of solitary waves where it is known as {\em{collective coordinate}} approach \cite{Scott}; it has since been used in the context of dissipative pattern forming systems \cite{GottwaldKramer04,MenonGottwald05,MenonGottwald07,MenonGottwald09,CoxGottwald06}. The method we propose makes explicit use of the functional form of the phases as suggested by numerical simulations. The parameters describing the functional form of the phases constitute the collective coordinates. For example, if observations reveal that the functional form of the solution is bell-shaped at all times, the collective coordinates might be the amplitude and width of a Gaussian. The temporal evolution of the full solution is then described by the temporal evolution of the collective coordinates, i.e. how the amplitude and the width of the Gaussians evolve in time. Of course, a specific assumed functional form is typically only an approximation of the actual solution. To eek out most of the assumed ansatz the collective coordinates are determined to optimally describe the solution. The most appropriate notion of optimality is to require that the error made by restricting the solution to be of the assumedz ansatz is minimised. Minimisation is achieved if the error is orthogonal to the subspace of the solutions spanned by the collective coordinates. This projection yields an evolution equation for the collective coordinates which allows to describe the actual solution at all times.\\

\begin{figure}
\begin{center}
\includegraphics[width=0.5\textwidth, height=0.25\textheight]{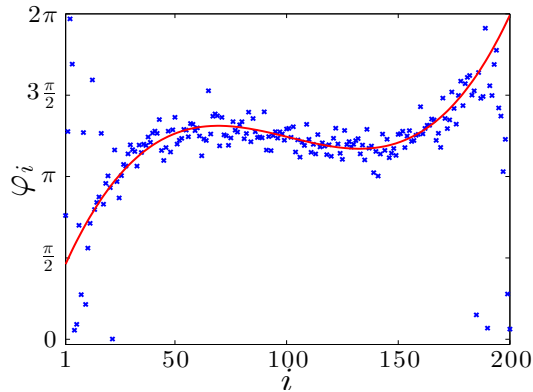}
\end{center}
%\caption{Snapshot of the phases $\varphi_j$ for $K=14.5$ for an $N=200$ obtained for the Kuramoto model with an Erd\H{o}s-R\'enyi topology and native frequencies $\omega=\xi^3$ with $\xi\sim{\mathcal{U}}[-1,1]$. The continuous line depicts a smooth cubic curve through $\omega$.}
\caption{Snapshot of the phases $\varphi_j$ for $K=9.5$ for an $N=200$ obtained for the Kuramoto model with an Erd\H{o}s-R\'enyi topology (with two nodes being connected with probability $p=0.05$) and native frequencies $\omega_i=\xi_i^3-0.3 \,\xi_i$ with $\xi_i\sim{\mathcal{U}}[-1,1]$. The continuous line depicts a smooth cubic function. The corresponding value of the order parameter is $\rbar=0.78$.}
\label{f.cubicomega}
\end{figure}

We now establish the method of collective coordinates for the Kuramoto model in detail. Without loss of generality we assume that the mean frequency is zero (unless stated otherwise). Let us assume that the nodes are labelled in order of increasing native frequencies, i.e. $i=1$ denotes the node with the most negative native frequency $\omega_1$ and $i=N$ denotes the node with the most positive native frequency $\omega_N$. In Figure~\ref{f.cubicomega} we show a snapshot of the phases $\varphi_j$ obtained by a numerical simulation of the Kuramoto model with an underlying Erd\H{o}s-R\'enyi topology with $N=200$ oscillators at a coupling strength $K=9.5$. The associated order parameter is $\rbar=0.78$ indicating a high level of synchronisation. The figure shows that the phases of oscillators with native frequencies of sufficiently small absolute value are frequency locked and correlate highly with the underlying native frequency distribution. This observation suggests that the phases of those frequency-locked oscillators may be approximated by
\begin{align}
\label{e.c1}
\varphi_i(t) = \alpha(t)\, \omega_i\; .
\end{align}
Oscillators with large absolute native frequencies which could not be entrained at a given coupling strength, do not obey this functional relationship but rather oscillate with their native frequencies. The ansatz (\ref{e.c1}) is trivially exact for $K=0$ with $\alpha(t) = t$. Furthermore, in the case of an all-to-all coupling the ansatz (\ref{e.c1}) can be formally motivated for large coupling strength as follows. The stationary Kuramoto model (\ref{e.kuramoto}) can be rewritten as $\omega_i = - K r \sin(\psi-\varphi_i)$ with $\psi$ being the mean phase \cite{Kuramoto}. Expanding $\varphi_i=\psi +\arcsin(\omega_i/(rK))$ in $1/K$ for large coupling strength yields up to first order $\varphi_i = \psi + \omega_i/(rK)$. Since the Kuramoto model (\ref{e.kuramoto}) is invariant under constant phase shifts we may set $\psi=0$ leading to our  ansatz (\ref{e.c1}).\\ Our method consists of assuming that the phases of the $N$ oscillators are approximately given by our ansatz (\ref{e.c1}). The time-dependent amplitude $\alpha(t)$ takes the role of a collective coordinate. Our goal is to find an evolution equation for $\alpha(t)$ and thereby reducing the $N$-dimensional Kuramoto model of phase oscillators to a one-dimensional ordinary differential equation for $\alpha(t)$ (in Section~\ref{sec-ATABimodal} we will see how to modify the approach to include more collective coordinates). We do so by requiring that the error 
\begin{align*}
\mathcal{E}_{\alpha} = \dot \alpha \omega_i - \omega_i - \frac{K}{N} \sum_{j=1}^Na_{ij}\sin(\alpha(\omega_j-\omega_i))
\end{align*}
made by restricting the solution to the subspace defined by the ansatz (\ref{e.c1}) is minimised. This is achieved by assuring that the error $\mathcal{E}_{\alpha}$ is orthogonal to the restricted subspace spanned by (\ref{e.c1}). We therefore require that the error $\mathcal{E}_{\alpha}$ is orthogonal to the tangent space of the solution manifold (\ref{e.c1}) which is spanned by $\partial \varphi_i/\partial \alpha = \omega_i$. Projecting the error onto the restricted subspace spanned by (\ref{e.c1}) yields the desired evolution equation for $\alpha$
\begin{align}
\dot \alpha = 1 + \frac{K}{\Sigma^2}\frac{1}{N^2} \sum_{i=1}^N\omega_i \sum_{j=1}^N a_{ij}\sin(\alpha(\omega_j-\omega_i))\; ,
\label{e.ccN}
\end{align}
with 
\begin{align}
\Sigma^2 = \frac{1}{N}\sum_{j=1}^N\omega_j^2\; .
\label{e.sigma2}
\end{align}
Solutions $\alpha^\star$ solving (\ref{e.ccN}) with $\dot \alpha = \Omega$ correspond to phase-locked solutions rotating uniformly with frequency $\Omega$ and phases $\varphi_j = \alpha^\star\omega_j + \Omega t$. The existence of such solutions corresponds to a synchronised state. The advantage of this approach is that it allows to study the onset of synchronisation of the $N$-dimensional network by analysing a one-dimensional problem and furthermore that it allows to study synchronisation for finite network size $N$.\\

In the limit $N\to \infty$ we can simplify the expressions by introducing the frequency distribution $g(\omega)$ and the variance of the frequencies $\sigma_\omega^2=\lim_{N\to\infty}\Sigma^2$. We obtain in an all-to-all coupling network with $a_{ij}=1$ for all $i,j$
\begin{align}
\dot \alpha = 1 + \frac{K}{\sigma_\omega^2}\int \omega g(\omega)\int \sin(\alpha(\eta-\omega)) g(\eta) d\eta\, d\omega\; .
\label{e.cc}
\end{align}

The order parameter $\rbar$ restricted to solutions $\varphi_j(t)=\alpha(t) \, \omega_j$ is introduced as
\begin{align}
\hat re^{i\psi}=\frac{1}{N}\sum_{j=1}^Ne^{i\alpha \omega_j}\; .
\end{align}
%where the mean phase $\psi$ can be set to zero. 
In the limit $N\to \infty$ the real part yields
\begin{align}
\hat r=\int \cos(\alpha \omega)g(\omega) \, d\omega\, ,
\label{e.rhatgeneral}
\end{align}
where we used that our ansatz (\ref{e.c1}) implies for the mean phase $\psi=0$.

We remark that this approach is not restricted to all-to-all network topologies. For example,  in an Erd\H{o}s-R\'enyi network, where nodes are connected independently with probability $p$ and where degrees $d_j$ are Poisson-distributed with mean degree $d=pN$, the inner sum in (\ref{e.ccN}) can be evaluated as a sum of (on average) $d$ random variables $\eta_j\sim g(\omega)$ with
\begin{align*}
\lim_{N\to \infty} \sum_{j=1}^Na_{ij}\sin(\alpha(\omega_j-\omega_i)) = d \int \sin(\alpha(\eta-\omega_i))g(\eta)d\eta\; .
\end{align*}
The evolution equation for $\alpha(t)$ is then evaluated in the limit $N\to \infty$ as
\begin{align}
\dot \alpha = 1 + p\frac{K}{\sigma_\omega^2}\int \omega g(\omega)\int \sin(\alpha(\eta-\omega)) g(\eta) d\eta\, d\omega\; .
\label{e.ccER}
\end{align}

In the next Section we will employ our framework to study the synchronisation properties of all-to-all coupling networks for several frequency distributions $g(\omega)$.

%

%%%%%%%%%%%%%%%%%%%%%%%%%%%%%%%%%%%%%%%
\section{Examples}
\label{sec-numerics}
%%%%%%%%%%%%%%%%%%%%%%%%%%%%%%%%%%%%%%%
%\subsection{All-to-all coupling}
%\label{sec-alltoall}
We now set out to illustrate the capabilities of the collective coordinate approach to describe the synchronisation behaviour of phase oscillators in a Kuramoto model with an all-to-all coupling topology. We do so by determining the steady state solution $\alpha$ and the order parameter $\hat r$, in the case of finite $N$ as well as in the thermodynamic limit of $N\to \infty$, for three different distributions of the native frequencies: uniform distribution, normal distribution and bimodal distribution. The results from the collective coordinate approach are then compared with results from direct numerical simulations of the corresponding Kuramoto model (\ref{e.kuramoto}).

Rather than performing averages over realisations of the native frequencies according to the respective distributions we will perform the calculations for the collective coordinate approach by choosing $N$ values of the native frequencies such that the probability of a random draw of a native frequency to fall in the interval $(\omega_{i},\omega_{i+1})$  is equal for all values of $i$. 

\subsection{Uniform distribution of native frequencies} 
\label{sec-ATAUniform}
In a first suite of experiments we consider native frequencies which are distributed uniformly on the interval $[-1,1]$ with distribution
\begin{align}
g(\omega)=0.5 \;.
\end{align}
Dividing the compact support of the frequency distribution $[-1,1]$ into $N-1$ intervals of equal measure, i.e. $\omega_i = 2(i - (N + 1)/2)/(N- 1))$ for $i=1,\cdots,N$, the evolution equation (\ref{e.ccN}) for $\alpha$ for finite $N$ is readily evaluated as
\begin{align}
\nonumber
%\dot \alpha &= 1 + \frac{3}{2}K\frac{1}{N^2(N+1)}\frac{1}{\sin^2 \frac{\alpha}{N-1} }\\
\dot \alpha &= 1 + \frac{3}{2}K\frac{1}{N^2(N+1)}\csc^2 \frac{\alpha}{N-1} \\
&\times \left(
 \cot \frac{\alpha}{N-1} \left(\cos \frac{2N\alpha}{N-1} - 1\right) + N\sin \frac{2N\alpha}{N-1}\right)\; ,
\label{e.ccN_uniform}
\end{align}
with 
\begin{align*}
\Sigma^2 = \frac{N+1}{3(N-1)}\, .
\end{align*}

%We omit the rather cumbersome and lengthy expression here, but present the expression 
In the thermodynamic limit this simplifies to
\begin{align}
\dot \alpha = 1 +\frac{K}{\sigma_\omega^2}\frac{\sin \alpha}{\alpha} \, \frac{\alpha\cos\alpha-\sin \alpha}{\alpha^2}\, ,
\label{e.alphadotuniform}
\end{align}
with $\sigma_\omega^2=\lim_{N\to \infty}\Sigma^2=1/3$.\\
The expression (\ref{e.rhatgeneral}) for the order parameter simplifies in the thermodynamic limit to 
\begin{align}
\hat r=\frac{\sin(\alpha)}{\alpha}\; .
\label{e.rhat_uniform}
\end{align}

\begin{figure}
\begin{center}
\includegraphics[width=0.5\textwidth, height=0.2\textheight]{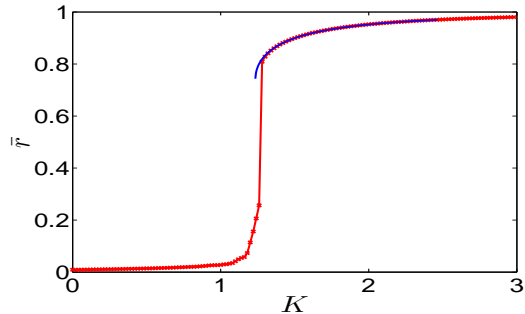}
\end{center}
\caption{Order parameter $\bar r$ as a function of the coupling strength $K$ for a network with uniformly distributed native frequencies. Depicted are results from a direct numerical integration of the Kuramoto model (\ref{e.kuramoto}) with $N=10,000$ (crosses, online red) and from the collective coordinate approach (\ref{e.rhat_uniform}) (continuous line, online blue).}
\label{f.uniform_CollCoo}
\end{figure}

\begin{figure}
\begin{center}
\includegraphics[width=0.5\textwidth, height=0.2\textheight]{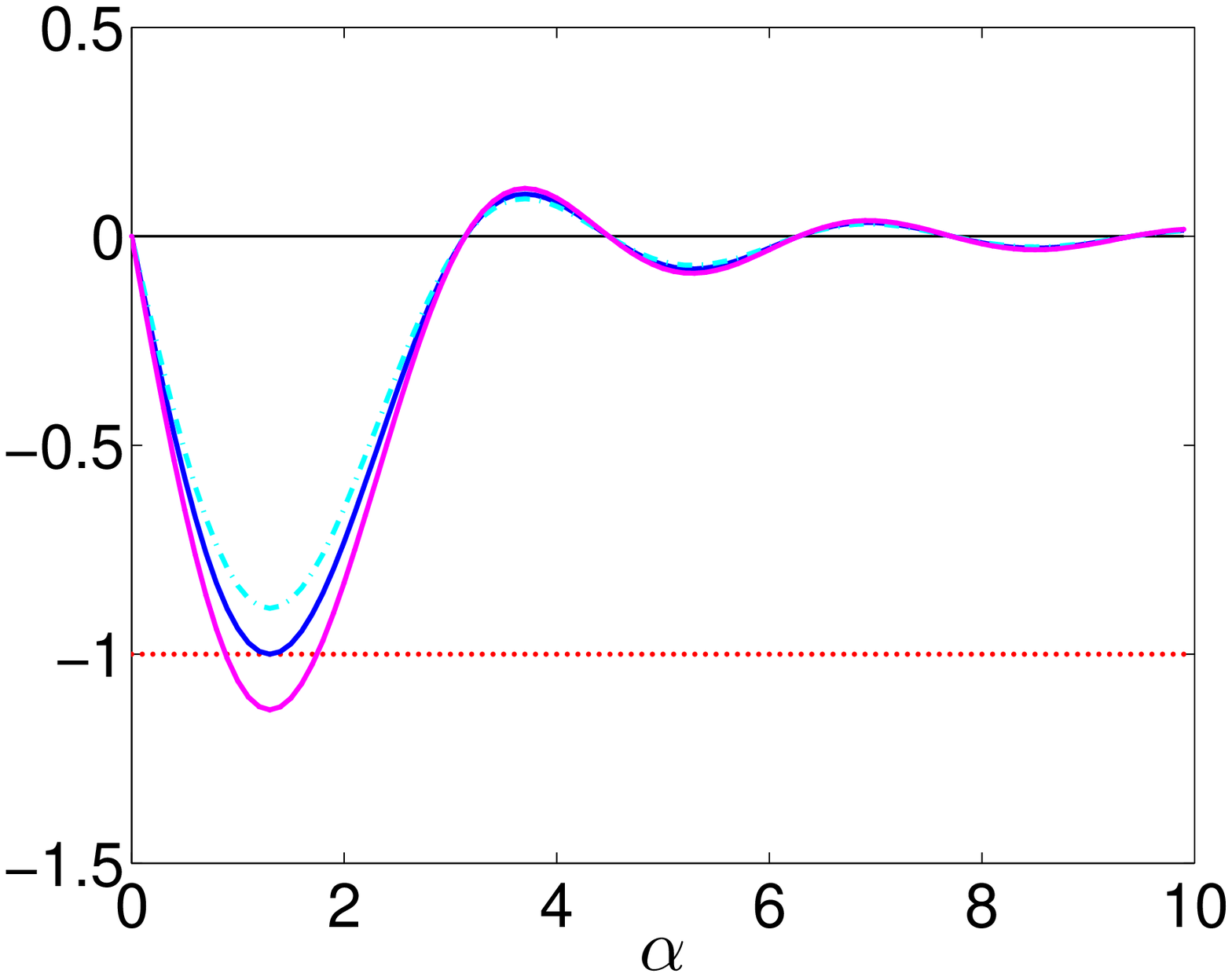}
\includegraphics[width=0.5\textwidth, height=0.2\textheight]{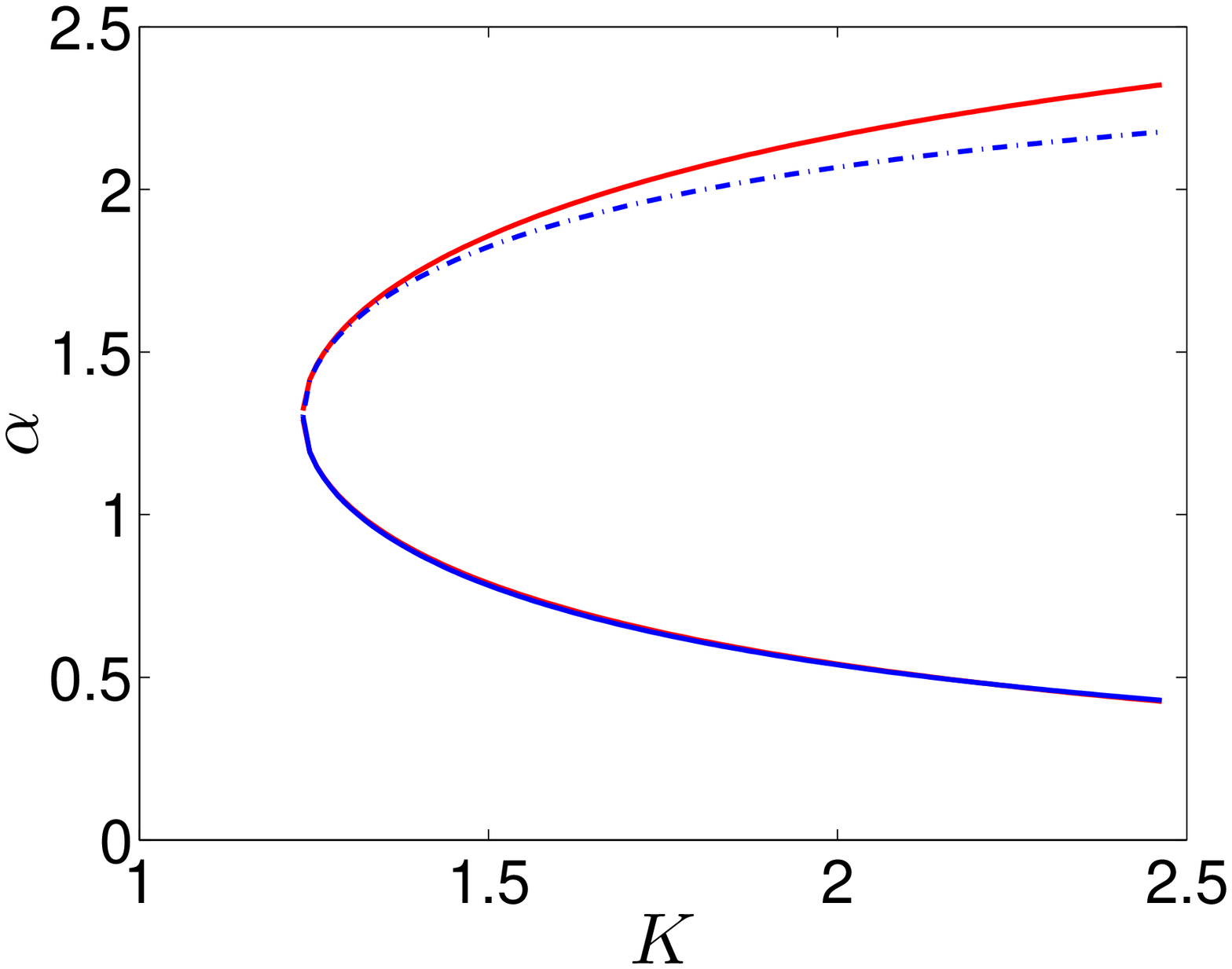}
\end{center}
\caption{Top: Plot of the term $\textstyle{\sum_i^N\omega_i^2}$ (dotted horizontal line, online red) and the term proportional to the coupling strength $K$ in (\ref{e.ccN}) as a function of the collective coordinate $\alpha$ for a network with uniformly distributed native frequencies. Intersections denote stationary solutions of (\ref{e.ccN}). Depicted are the subcritical case at $K=1.1$ (dashed curve, online cyan), the critical case $K=K_c = 1.234$ (continuous curve, online blue) and the supercritical case with $K=1.4$ (continuous line with circles, online magenta) for $N=1000$. At criticality we find $\alpha_c=1.303$.
Bottom: The stable and unstable stationary solutions $\alpha$ as a function of the coupling strength $K$ as calculated from the collective coordinate approach (\ref{e.ccN}) (continuous lines) and from the approximation (\ref{e.alpha_SN}) (dashed lines). The two approximations are hardly distinguishable on the lower stable branch.}
\label{f.uniform_SN}
\end{figure}

In Figure~\ref{f.uniform_CollCoo} we show the order parameter $\rbar$ as a function of the coupling strength $K$ obtained from a long time integration of the full Kuramoto model (\ref{e.kuramoto}). The onset of synchronisation appears to be hard (see for example, \citet{Pazo05}), i.e. there exists a non-zero value of the order parameter at the critical coupling strength $K_c$. The collective coordinate approach captures this very well as shown in Figure~\ref{f.uniform_CollCoo}. Figure~\ref{f.uniform_SN} shows that within the framework of collective coordinates the hard onset of synchronisation is described as a saddle node bifurcation~\cite{PikovskyEtAl}: for $K>K_c=1.234$ a pair of stationary solutions $\varphi_j=\alpha \omega_j$ (a smaller stable and a larger unstable one) exist; at criticality the two solutions collide in a saddle node bifurcation at $\alpha=\alpha_c\approx1.303$, and there are no stationary solutions for $K<K_c$. Evaluating the right-hand-side of (\ref{e.alphadotuniform}) around the critical value $\alpha_c$ yields as an approximation of the stable and unstable stationary solutions $\alpha_{\rm{s,u}}$ close to criticality 
\begin{align}
\alpha_{\rm{s,u}} = \alpha_c \pm m \sqrt{1-\frac{K_c}{K}}\; ,
\label{e.alpha_SN}
\end{align}
with the critical coupling strength $K_c$ and $m=\sqrt{0.270/0.177}$. Figure~\ref{f.uniform_SN} shows a numerical evaluation of the stationary solutions $\alpha$ of (\ref{e.c1}) as well as the approximate solutions (\ref{e.alpha_SN}). Note that the stable stationary solution is well approximated for a large range of coupling strengths $K$ even far away from criticality.\\ We now analyse the order parameter $\hat r$ as given by (\ref{e.rhat_uniform}). Figure~\ref{f.uniform_CollCoo} shows the order parameter as a function of the coupling strength obtained from a numerical simulation of a large network with $N=10,000$ nodes simulating the Kuramoto model (\ref{e.kuramoto}), and as calculated within the collective coordinate framework using (\ref{e.rhat_uniform}). The critical coupling strengths for the full Kuramoto model with $N=10,000$ is $K_c=1.279$ which is close to the exact analytical result for the thermodynamic limit with $K_c=4/\pi\approx 1.273$ \cite{Kuramoto,Pazo05,UmEtAl14}. This is well approximated by our simple model with an error of $3\%$. The non-zero order parameter at the hard transition, which is $r_c=\pi/4\approx 0.785$ in the thermodynamic limit \cite{Kuramoto,Pazo05,UmEtAl14}, is estimated as $r_c=0.744$ within the collective coordinate approach implying a $5\%$ error. Note that the order parameter is extremely well approximated for large values of the coupling strength. This is not surprising since, as pointed out in Section~\ref{sec-coll}, the collective coordinate ansatz (\ref{e.c1}) is consistent with an expansion of the stationary solution in $1/K$ for all-to-all coupling networks.\\

\begin{figure}
\begin{center}
\includegraphics[width=0.5\textwidth, height=0.2\textheight]{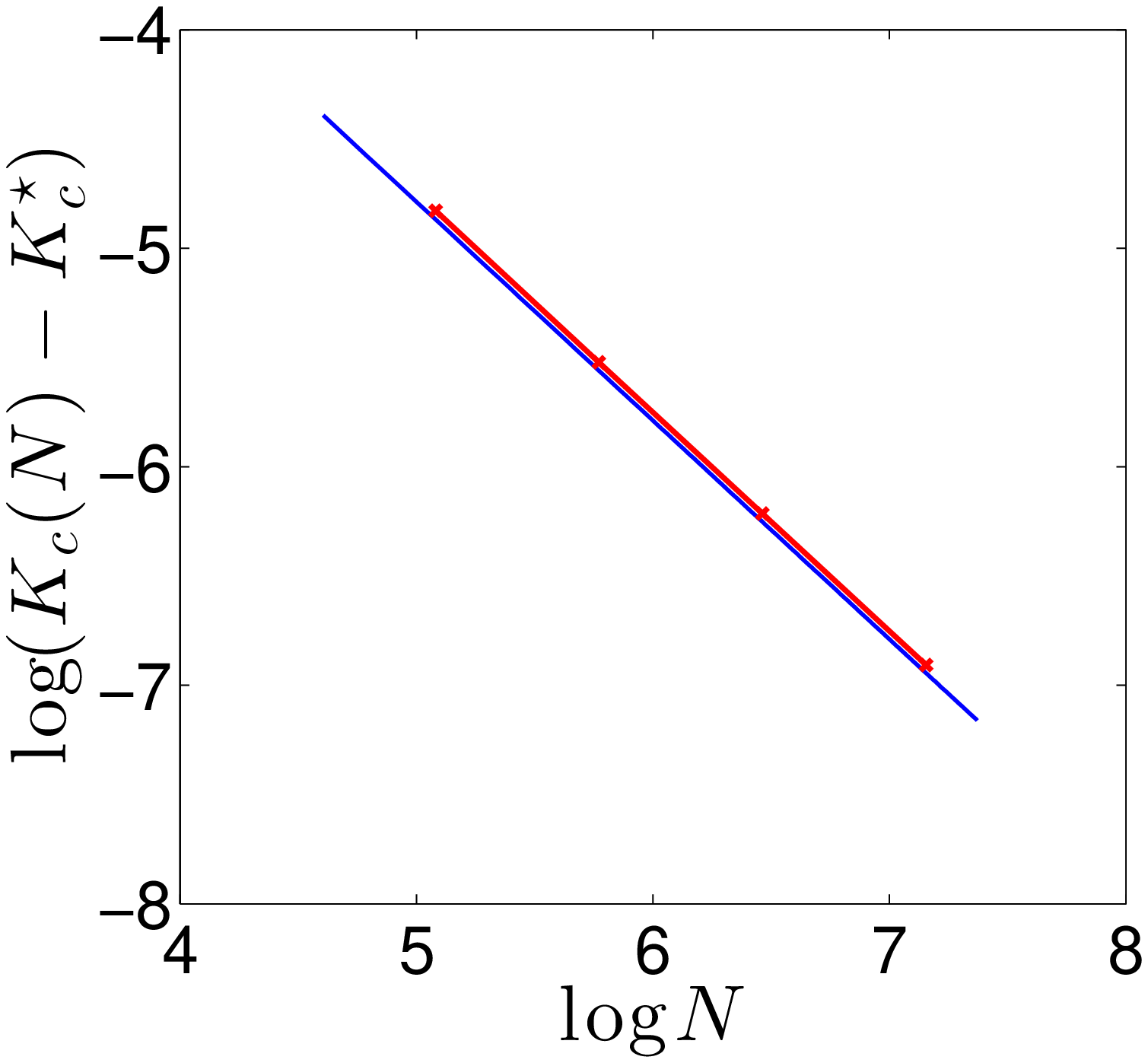}
\end{center}
\caption{Scaling of the critical coupling strength $K_c$ as a function of the network size $N$ for a network with uniformly distributed native frequencies. Depicted are results from a direct numerical integration of the Kuramoto model (\ref{e.kuramoto}) (crosses, online red) and from the collective coordinate approach (continuous line, online blue). The two lines have slopes of $1$.}
\label{f.uniform_finitesize}
\end{figure}

A particular advantage of our approach is that it allows us to study the finite size scaling of synchronisation behaviour \cite{Pazo05,HildebrandtEtAl07,HongEtAl07,Tang11}. In Figure~\ref{f.uniform_finitesize} we show a comparison of the critical coupling strength $K_c(N)$ as calculated via our collective coordinate approach for variable network sizes $N$ and results from direct simulations of the Kuramoto model (\ref{e.kuramoto}). The difficulty is determining the critical coupling strength $K_c$ in finite size networks is that the order parameter has fluctuations of the order $1/\sqrt{N}$ which confounds the onset. As a proxy for the critical coupling strength we record for each value of $N$ the smallest value of the coupling strength $K$ such that $\bar r>0.8$. We have also used the criterion whereby the critical coupling is determined as the coupling strength when the minimal value of the order parameter $r(t)$ over some sufficiently long time window changes from values close to zero to values significantly above zero \cite{TemirbayevEtAl12}. This method yields very pronounced onsets, but is not able to detect global synchronisation in the case when it is preceded by partial synchronisation. We therefore present only results obtained using the first method. In the case of uniformly distributed native frequencies, however, both methods yield the same results.\\ Linear regression suggests a scaling $K_c(N)-K_c^\star\sim N$ where we estimate the critical coupling strength in the thermodynamic limit $K_c^\star$ as $K_c^\star = 1.279$ for the full Kuramoto model and $K_c^\star = 1.234$ for the collective coordinate approach.\\

\begin{figure}
\begin{center}
\includegraphics[width=0.48\textwidth, height=0.25\textheight]{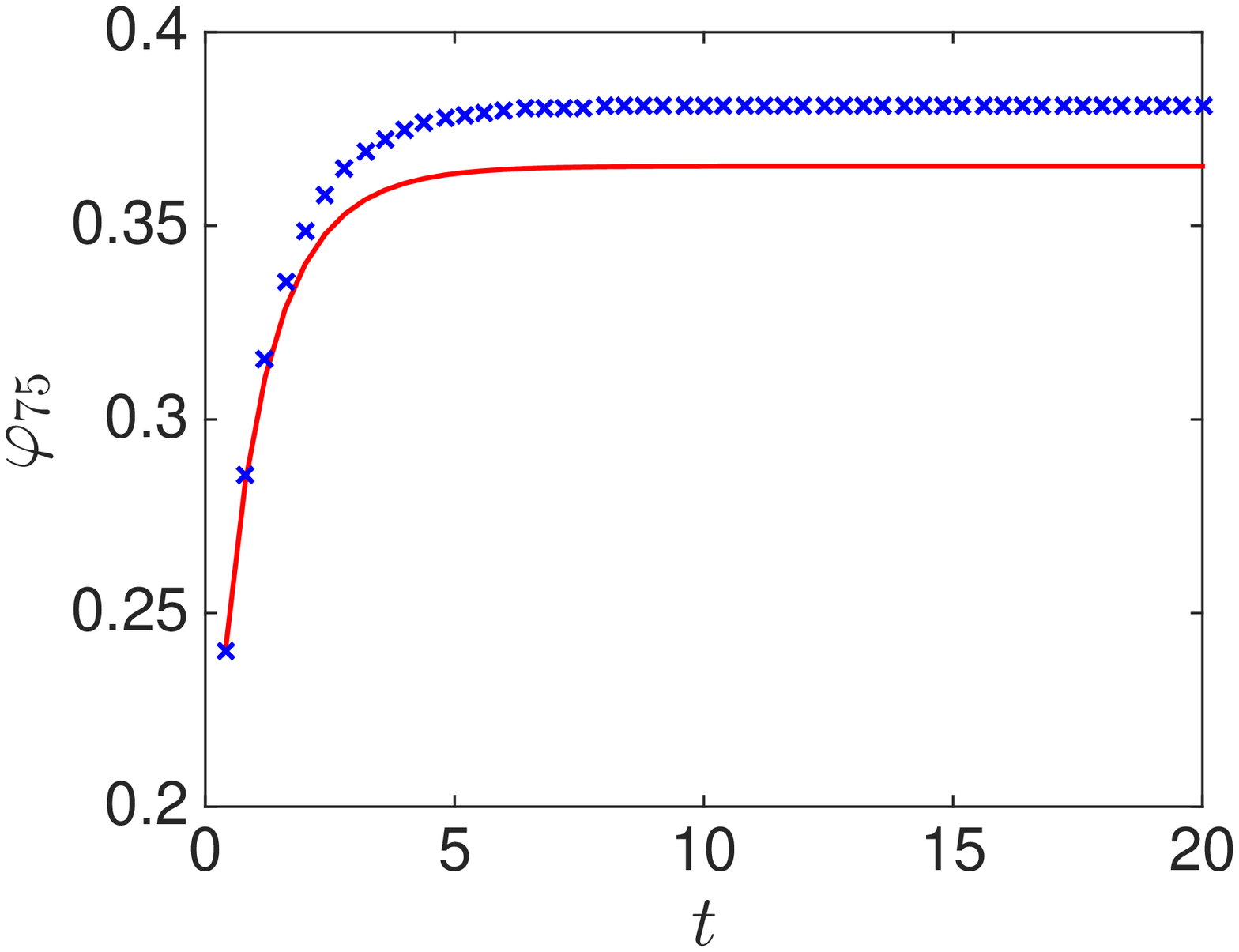}
\includegraphics[width=0.48\textwidth, height=0.25\textheight]{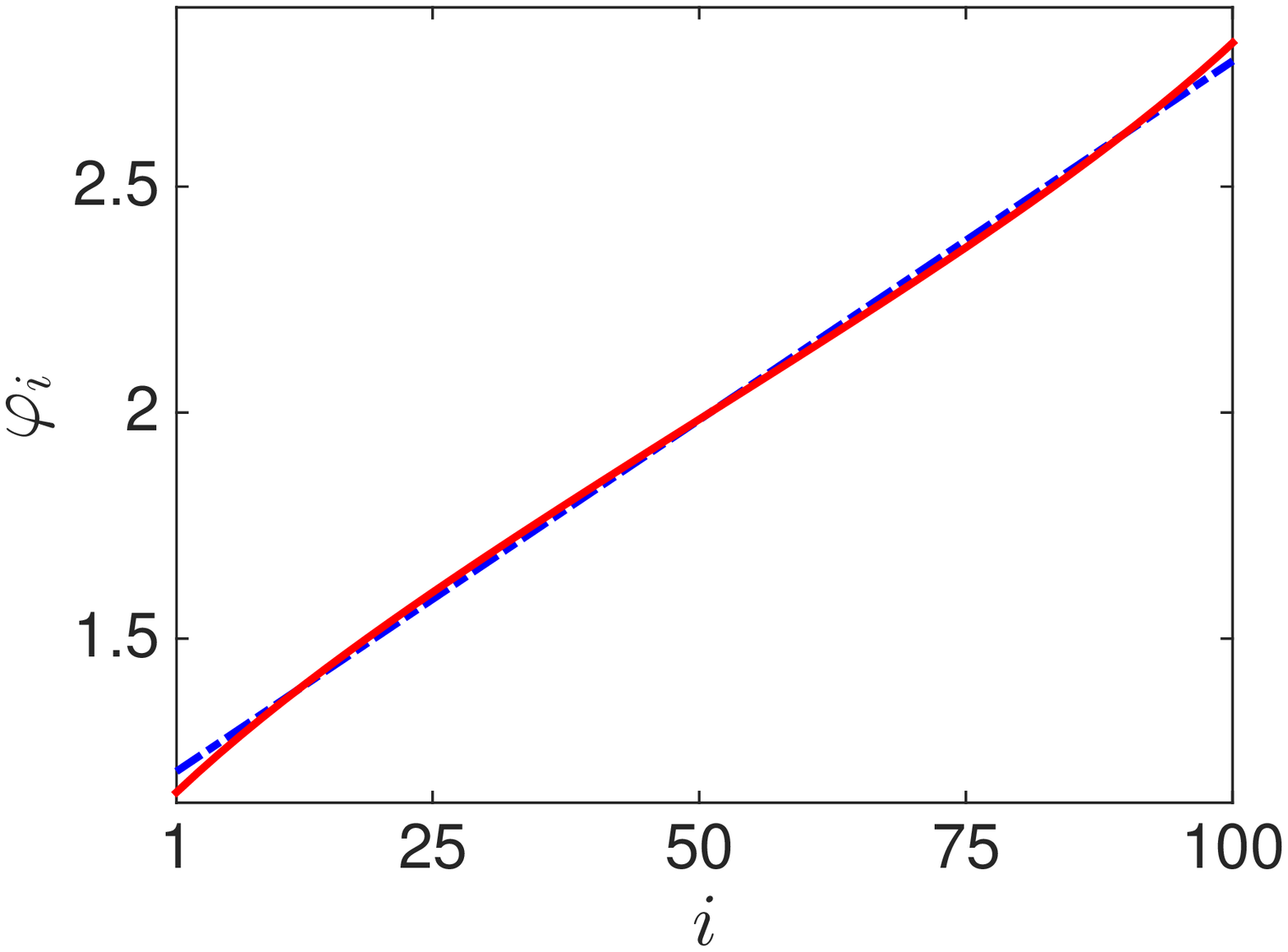}
\end{center}
\caption{Phases $\varphi(t)$ calculated from simulations of the full Kuramoto model (\ref{e.kuramoto}) (continuous lines, online red) and from the corresponding $1$-dimensional system (\ref{e.ccN}) for the collective coordinate with $\varphi_i = \alpha(t)\omega_i$ (crosses, online blue) for a network of $N=100$ oscillators with uniformly distributed native frequencies  at coupling strength $K=1.5$. Top: Temporal evolution of $\varphi_{75}(t)$ for initial conditions $\varphi_i(0) = \alpha_0 \omega_i$ with $\alpha_0 = 0.5$. Bottom: Snapshot of the phases $\varphi_i(T)$ at time $T=20$.}
\label{f.uniform_TE1}
\end{figure}

Besides being able to describe the collective behaviour of oscillators and the onset of synchronisation, we now show that the collective coordinate approach also captures the temporal evolution of individual oscillators through the evolution equation (\ref{e.ccN}) or its equivalent formulation (\ref{e.ccN_uniform}) for uniformly distributed native frequencies. For sufficiently small coupling strengths $K$, where the oscillators only weakly interact, both models produce indistinguishable trajectories with phases growing linear in time 
%according to $\varphi_j = \varphi_j(0)  + \alpha_0 \omega_j t$ 
 (not shown). Figure~\ref{f.uniform_TE1} shows a comparison of actual trajectories for a network with $N=101$ oscillators at coupling strength $K=1.5>K_c$ where the collective coordinate approach describes the order parameter $\bar r\approx 0.9$ very well (cf. Figure~\ref{f.uniform_CollCoo}). We show a comparison of the phase of the $75$th oscillator $\varphi_{75}$ with native frequency $\omega_{75}=0.48$ is obtained by solving the full Kuramoto model (\ref{e.kuramoto}) and by solving (\ref{e.ccN_uniform}) for the collective coordinate approach (\ref{e.c1}). If the initial conditions are chosen to satisfy $\varphi_j(0)=\alpha_0\omega_j$ with the initial condition $\alpha(0)=\alpha_0$ not too far away from its equilibrium solution, the two trajectories are reasonably close (top panel). This correspondence of the time evolution of the solutions of the full Kuramoto model and the collective coordinate approach is destroyed for initial conditions which are too far from the asymptotic state, i.e. if $\alpha_0$ is chosen too large. Their asymptotic state, however, will be close  and both systems will evolve to the same fix point, implying that the order parameter $\rbar$ will be close for the two systems. Similarly, if the initial conditions $\varphi_j(0)$ of the Kuramoto model are distributed around the initial condition the asymptotic state and therefore differs from the initial condition $\varphi_j(0)$ implied by the collective coordinate ansatz (\ref{e.c1}), the asymptotic temporal evolution of the full Kuramoto model and the reduced collective coordinate system are close (not shown). This is consistent with the previous observation that the order parameters $\rbar$ are close for the respective systems, as shown in Figure~\ref{f.uniform_CollCoo}. We show a snapshot depicting the phases of all oscillators in the phase-locked state illustrating that the collective coordinate approach captures the dynamics of the full model. Deviations occur for the extreme oscillators with largest absolute value of the native frequencies.  As we have seen in Figure~\ref{f.uniform_CollCoo} the collective coordinate approach predicts the onset of synchronisation for smaller values of $K$ than observed for the actual Kuramoto model. For coupling strength where the order parameter significantly differs between the reduced model and the full model, there is of course, also no correspondence between the temporal evolution of the phases nor their asymptotic dynamics. We remark that we obtain similar results for networks differing in several orders of magnitude in size. For small networks of, for example, size $N=20$, the phases are very well recovered if the native frequencies are chosen such that they divide the interval $[-1,1]$ into equiprobable partitions. For a particular random draw from the uniform distribution, the phases and their asymptotic states may differ though, in particular for oscillators with large absolute native frequencies. This discrepancy can be alleviated for the well-synchronised oscillators if averages over many realisations of native frequencies are taken. With increasing size of the networks, the difference between solutions obtained for random realisations of the native frequencies become smaller.\\


\subsection{Normal distribution of native frequencies}
\label{sec-ATAGaussian}
In a second suite of experiments, we consider native frequencies which are normally distributed with $\omega_i\sim{\mathcal{N}}(0,\sigma_\omega^2)$. The distribution is given by
\begin{align}
g(\omega)=\frac{1}{Z} \exp\left(-\frac{\omega^2}{2 \sigma_\omega^2}\right)\;,
\end{align}
with a normalisation constant $Z=\sqrt{2\pi\sigma_\omega^2}$. We use here $\sigma_\omega^2=0.1$. The evolution equation (\ref{e.ccN}) for $\alpha$ for finite $N$ can be evaluated for random draws of $\omega_i$, but we omit here the cumbersome expressions. In the thermodynamic limit the dynamic model for the collective coordinate (\ref{e.cc}) simplifies to
\begin{align}
\dot \alpha = 1 - K \alpha \exp(-\sigma_\omega^2\alpha^2)\; .
\label{e.alphadotgaussian}
\end{align}
The equation for the order parameter (\ref{e.rhatgeneral}) can be evaluated in the thermodynamic limit to
\begin{align}
\hat r=\exp\left(-\frac{\sigma_\omega^2 \alpha^2}{2}\right)\; .
\label{e.rhat_gaussian}
\end{align}

\begin{figure}
\begin{center}
\includegraphics[width=0.5\textwidth, height=0.2\textheight]{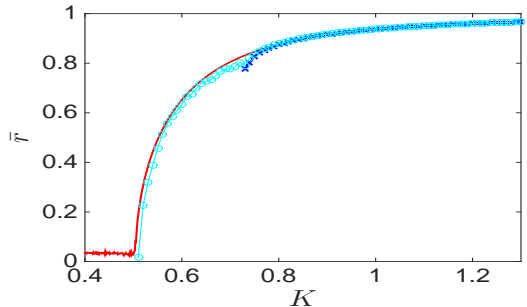}
\end{center}
\caption{Order parameter $\bar r$ as a function of the coupling strength $K$ for a network with normally distributed native frequencies. Depicted are results from a direct numerical integration of the Kuramoto model (\ref{e.kuramoto})  with $N=1000$ (continuous line, online red) and from the collective coordinate approach for global phase synchronisation (crosses, online blue) and for local phase synchronisation (open circles, online cyan). The curves coincide for sufficiently large values of the coupling strength $K$.}
%\caption{Order parameter $\bar r$ as a function of the coupling strength $K$ for a network with normally distributed native frequencies. Depicted are results from a direct numerical integration of the Kuramoto model (\ref{e.kuramoto})  with $N=1000$ (continuous line, online red) and from the collective coordinate approach for global phase synchronisation in the thermodynamic limit using (\ref{e.rhat_gaussian})  (crosses, online blue) and for local phase synchronisation with $N=1000$ (open circles, online green) using (\ref{e.ccN}).}
\label{f.Gaussian_CollCoo}
\end{figure}

\begin{figure}
\begin{center}
\includegraphics[width=0.5\textwidth, height=0.2\textheight]{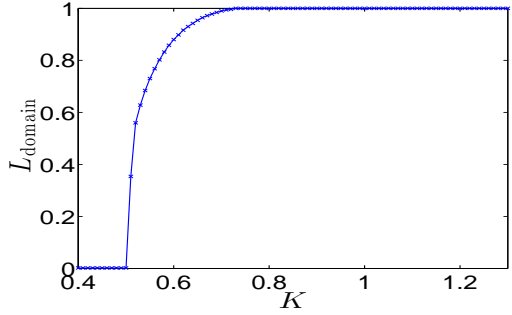}
\end{center}
\caption{Normalised length of phase synchronised domain as a function of the coupling strength for a network with normal native frequency distribution, calculated using the collective coordinate approach.}
\label{f.Gaussian_L}
\end{figure}

It is well known that in the case of unimodal frequency distributions, the onset of synchronisation is soft~\cite{Kuramoto,OsipovEtAl}. This is illustrated in Figure~\ref{f.Gaussian_CollCoo} where $\rbar$ is shown as a function of the coupling strength. At the so called ``Kuramoto coupling" $K=K_l$ the order parameter becomes non-zero and a few oscillators with native frequencies close to the mean frequency mutually synchronise; increasing the coupling strength allows more and more oscillators to synchronise, implying a continuous change of the order parameter $\rbar(K)$ as supposed to the hard transition in the case of uniformly distributed native frequencies described in the previous subsection. At some coupling strength $K=K_c$ global synchronisation sets in affecting all oscillators \cite{VerwoerdMason14}. \\ In the thermodynamic limit $N\to \infty$ the Kuramoto coupling can be approximated by $K_l = 2 /\pi g(0)\approx 0.505$ \cite{Kuramoto}. The transition to global synchronisation is not visible, however, by just looking at the order parameter $\rbar$ determined from numerical simulations of the full Kuramoto model (\ref{e.kuramoto}).\\ We will now show that the collective coordinate approach is able to describe both, the onset of global synchronisation at $K=K_c$ as well as the onset of local synchronisation at the ``Kuramoto coupling" $K=K_l$. The onset of global synchronisation can be calculated as before. In Figure~\ref{f.Gaussian_CollCoo} we show a result of the collective coordinate approach (\ref{e.rhat_gaussian}) which predicts the onset of global synchronisation at $K_c\approx 0.730$ with a non-zero value of $\bar r_c\approx0.779$.\\ By construction, the ansatz (\ref{e.c1}) cannot describe local synchronisation where only a subset of the $N$ phase oscillators are phase locked. We now modify the collective coordinate approach to allow for local synchronisation. We denote by $N_l$ the size of the mutually synchronised local group, consisting of those $N_l$ oscillators with frequencies closest to the mean frequency zero. Hence we restrict our solutions to obey
\begin{align}
\label{e.c2}
\varphi_j(t) = \alpha(t)\, \omega_j 
\quad {\rm for} \quad  
\frac{N-N_l}{2} \le j\le \frac{N+N_l}{2}\; .
%\frac{N-1}{2} - \frac{N_l}{2} \le j\le \frac{N-1}{2} + \frac{N_l}{2}\; ,
\end{align}
The evolution equation for the collective coordinate $\alpha(t)$ is again obtained by projecting the error made by the ansatz (\ref{e.c2}) onto the restricted subspace spanned by (\ref{e.c2}). We obtain
\begin{align}
\dot \alpha = 1 + \frac{K}{\Sigma_l^2}\frac{1}{N N_l} \sum_{i=N_{l1}}^{N_{l2}}\omega_i \sum_{j=N_{l1}}^{N_{l2}} \sin(\alpha(\omega_j-\omega_i))\; ,
\label{e.ccN2}
\end{align}
where the variance of the local group of frequencies is
\begin{align}
\Sigma_l^2 = \frac{1}{N_l}\sum_{j=N_{l1}}^{N_{l2}}\omega_j^2\; ,
\label{e.sigma2loc}
\end{align}
with $N_{l1}=\textstyle{\frac{N-N_l}{2}}$ and $N_{l2}=\textstyle{\frac{N+N_l}{2}}$
This is just the analogous formulation of (\ref{e.ccN}) for a group of oscillators, centred around $\omega_i=0$, of size $N_l$. Assuming that all those oscillators which can synchronise do so, the size of the locally synchronised group of oscillators $N_l$ can be determined as the maximal value of $N_l$ which supports stationary solutions of (\ref{e.ccN2}) for a given coupling strength $K$. Note that $N_l=N$ for $K\ge K_c$.\\ Figure~\ref{f.Gaussian_L} shows how the normalised domain length of the local synchronised cluster
\begin{align}
%L_{\rm{domain}} = \frac{N_l}{\frac{N-1}{2}}\; ,
L_{\rm{domain}} = \frac{N_l}{N}\; ,
\label{e.Ldomain}
\end{align}
increases from zero to $L_{\rm{domain}}>0$ at $K=K_l$ and then reaches $L_{\rm{domain}}=1$ at $K=K_c$ at which point global synchronisation sets in. The Kuramoto coupling, i.e. the smallest value of $K$ which gives rise to a non-zero value of $L_{\rm domain}$, is estimated for $N=1000$ by our approach as $K_l\approx 0.5$ corresponding very well with the numerically observed onset of local synchronisation. The asymptotic value is given by $K_l\approx 2/\pi g(0)\approx0.505$ \cite{Kuramoto}. It is pertinent to mention that in the case of uniformly distributed native frequencies, no stationary solutions $\alpha$ exist for any $N_l<N$, consistent with the absence of local synchronisation and the existence of a hard transition, as seen in Figure~\ref{f.uniform_CollCoo}.\\

\begin{figure}
\begin{center}
\includegraphics[width=0.5\textwidth, height=0.2\textheight]{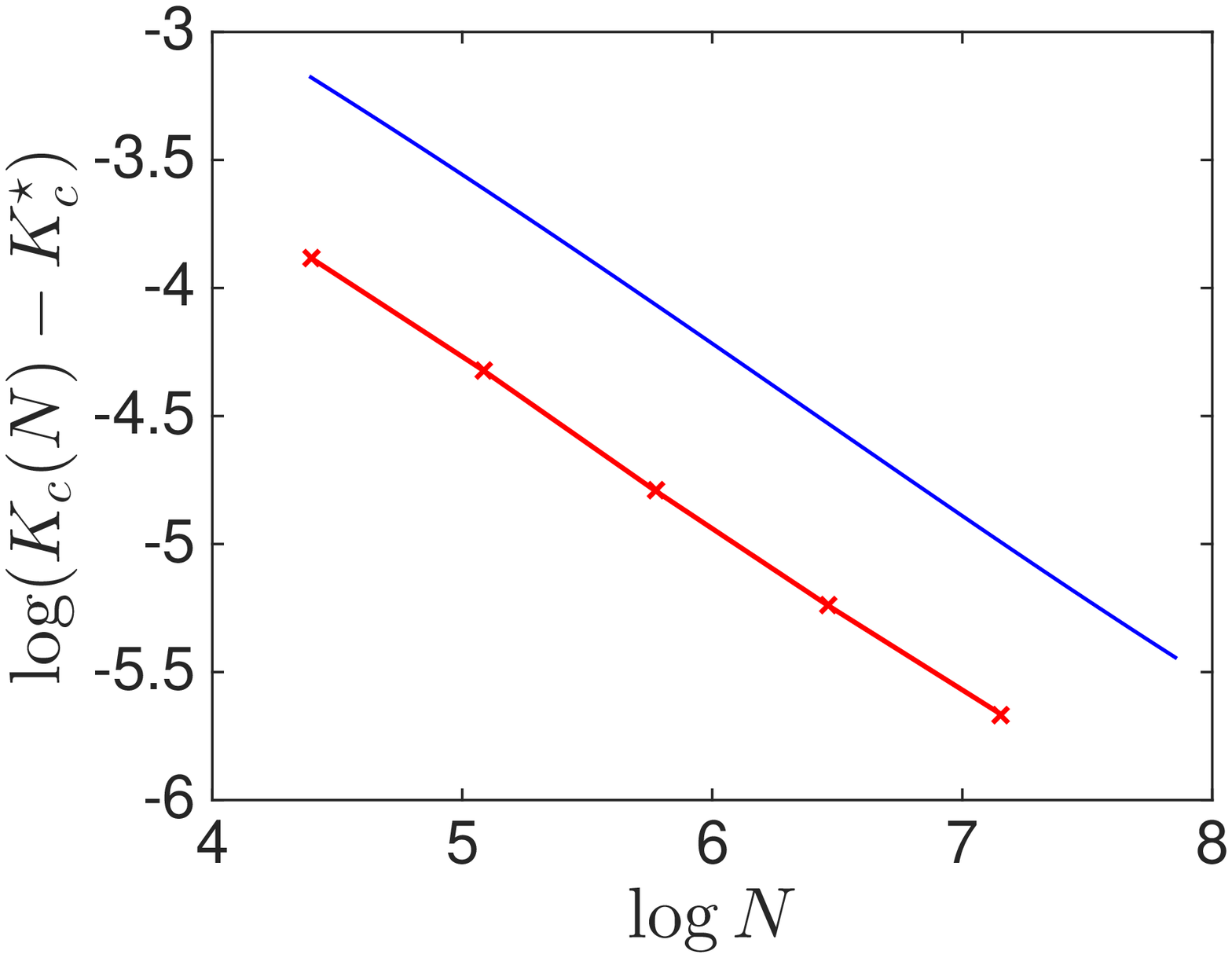}
\end{center}
\caption{Scaling of the critical coupling strength $K_c$ as a function of the network size $N$ for a network with normally distributed native frequencies. Depicted are results from a direct numerical integration of the Kuramoto model (\ref{e.kuramoto}) (crosses, online red) and from the collective coordinate approach (continuous line, online blue). The two lines have slopes of $0.664$ suggesting a scaling with $2/3$.}
\label{f.gaussian_finitesize}
\end{figure}

In Figure~\ref{f.gaussian_finitesize} we illustrate again that the collective coordinate approach can be used to study finite size scaling. For normally distributed native frequencies the numerics suggest a finite size scaling of $K_c(N)-K_c^\star(N) \sim N^{2/3}$.\\

\begin{figure}
\begin{center}
\includegraphics[width=0.48\textwidth, height=0.25\textheight]{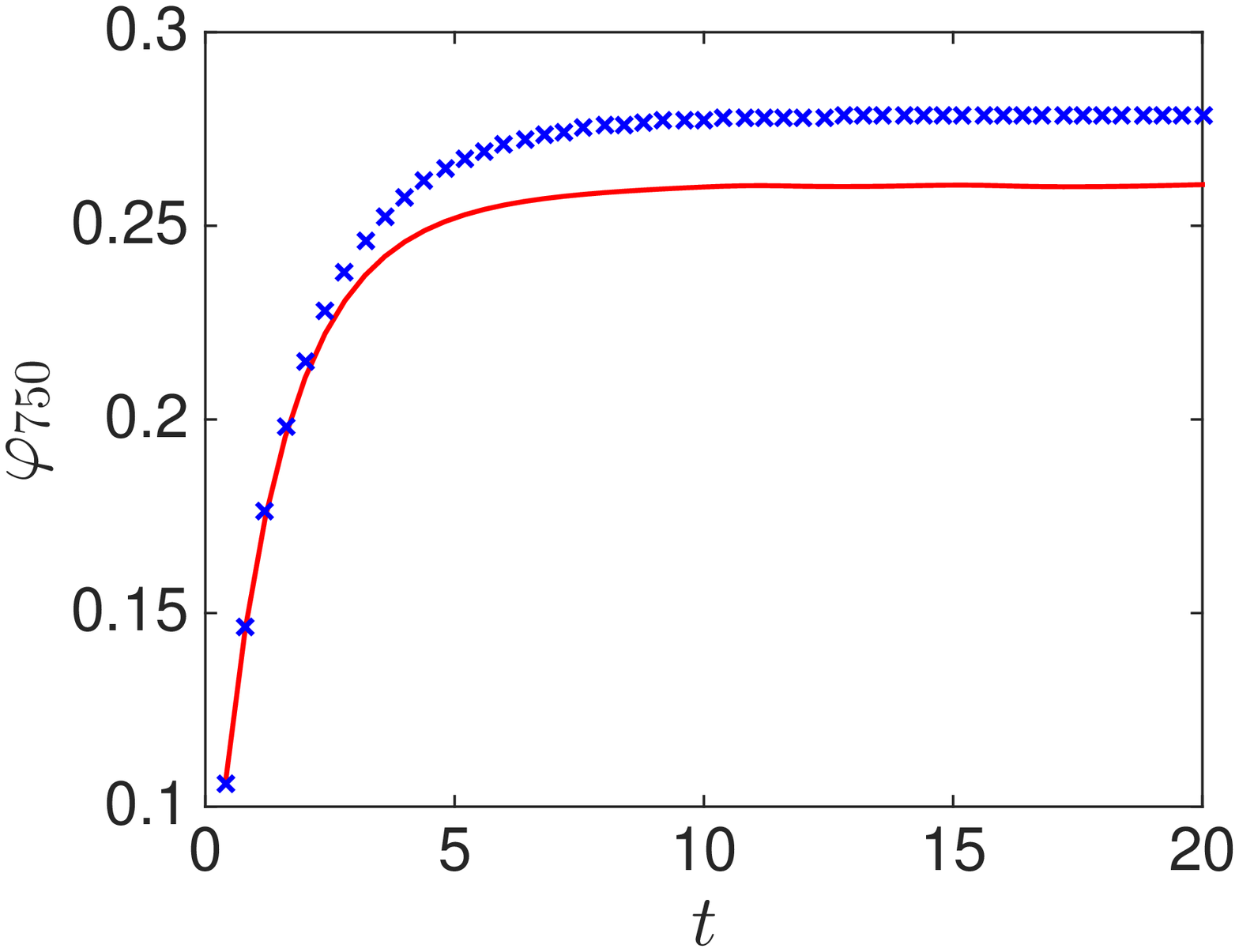}
\includegraphics[width=0.48\textwidth, height=0.25\textheight]{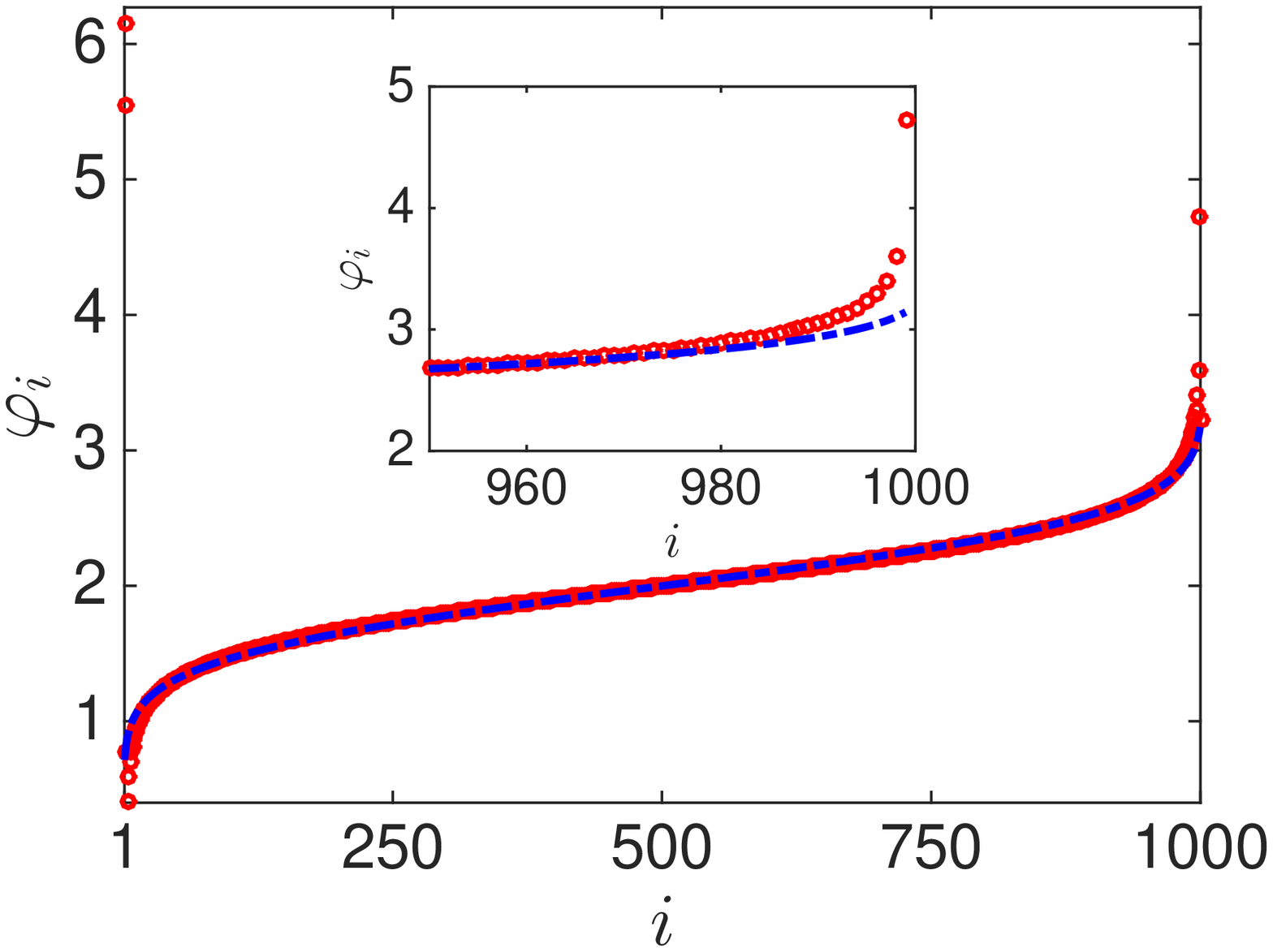}
\end{center}
\caption{Phases $\varphi(t)$ calculated from simulations of the full Kuramoto model (\ref{e.kuramoto}) (continuous lines or open circles, online red) and from the corresponding $1$-dimensional system (\ref{e.ccN}) for the collective coordinate with $\varphi_i = \alpha(t)\omega_i$ (crosses, online blue) for a network of $N=1000$ oscillators with normally distributed native frequencies at coupling strength $K=0.9$ corresponding to global synchronisation. Top: Temporal evolution of $\varphi_{750}(t)$ for initial conditions $\varphi_i(0) = \alpha_0 \omega_i$ with $\alpha_0 = 0.5$. The native frequency is $\omega_{750}=0.212$. Bottom: Snapshot of the phases $\varphi_i(T)$ at time $T=20$.}
%Open circles (online red) depict solution of the Kuramoto model (\ref{e.kuramoto}).
\label{f.Gaussian_TE1}
\end{figure}

\begin{figure}
\begin{center}
\includegraphics[width=0.48\textwidth, height=0.25\textheight]{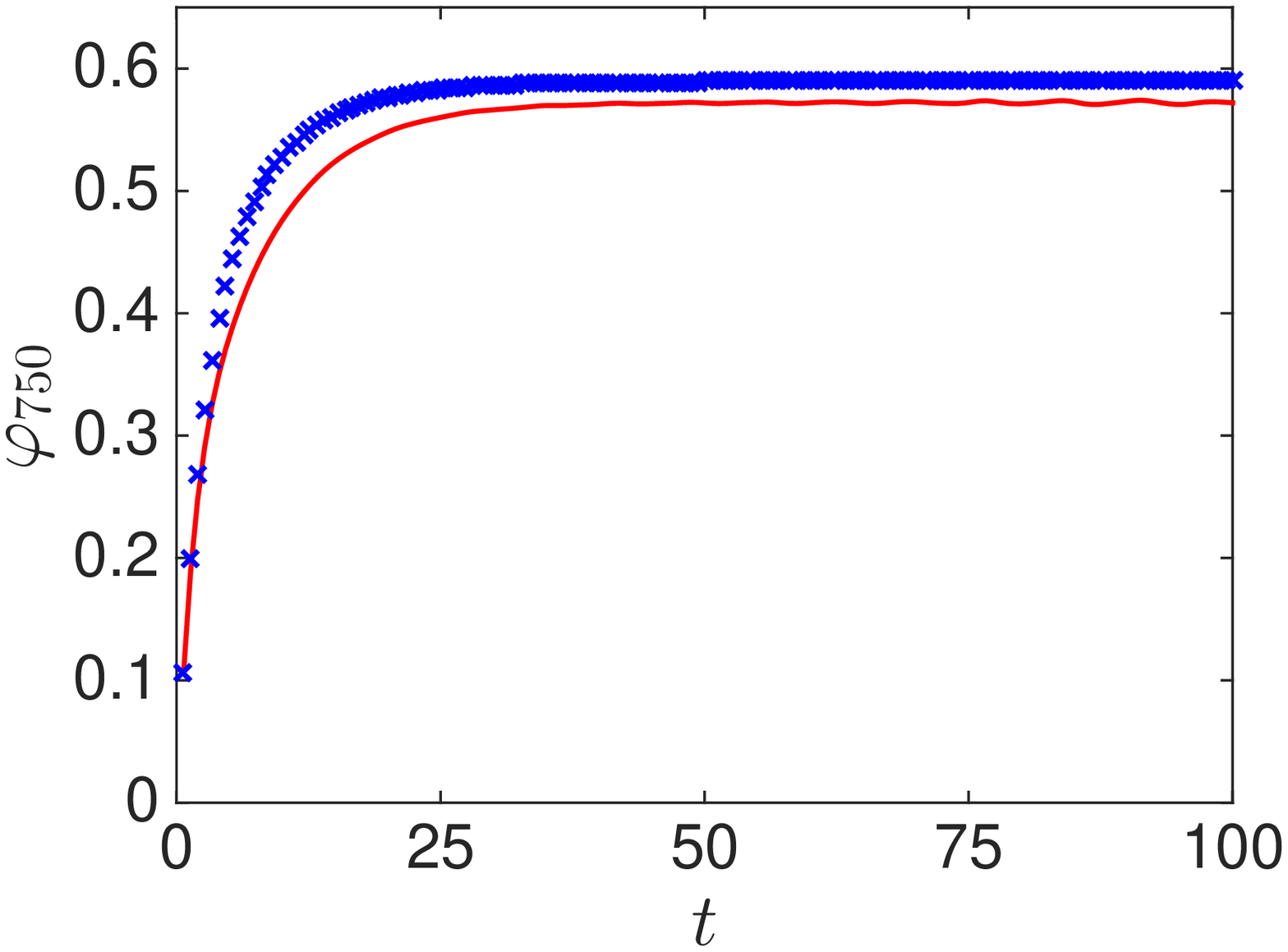}
\includegraphics[width=0.48\textwidth, height=0.25\textheight]{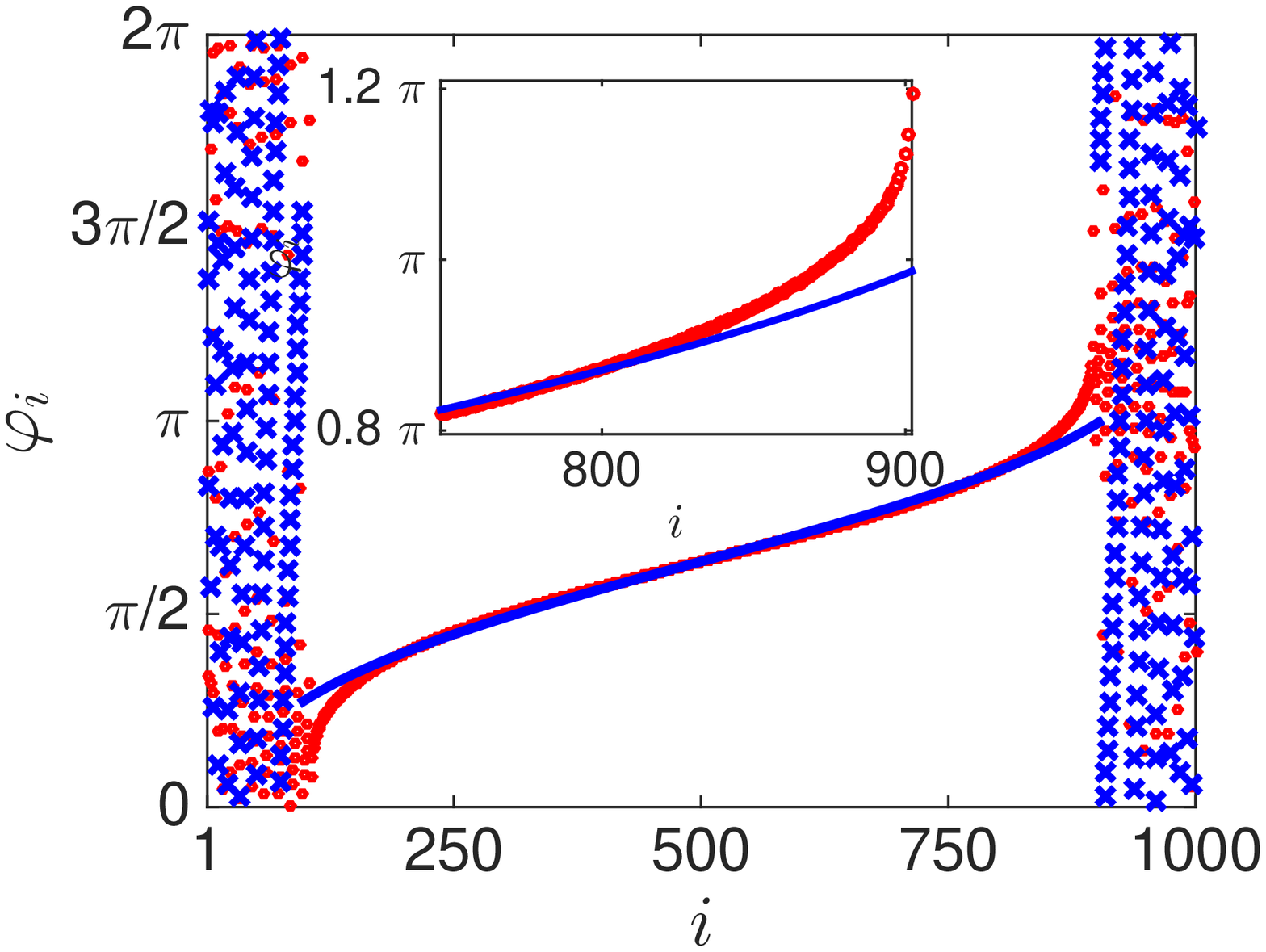}
\end{center}
\caption{Phases $\varphi(t)$ calculated from simulations of the full Kuramoto model (\ref{e.kuramoto}) (continuous lines or open circles, online red) and from the corresponding $1$-dimensional system (\ref{e.ccN2}) for the collective coordinate with $\varphi_i = \alpha(t)\omega_i$ (crosses or dashed line, online blue) for a network of $N=1000$ oscillators with normally distributed native frequencies at coupling strength $K=0.6$ with $N_l = 805$ corresponding to local synchronisation. Top: Temporal evolution of $\varphi_{750}(t)$ for initial conditions $\varphi_i(0) = \alpha_0 \omega_i$ with $\alpha_0 = 0.5$. The native frequency is $\omega_{750}=0.212$. Bottom: Snapshot of the phases $\varphi_i(T)$ at time $T=20$.}
\label{f.Gaussian_TE2}
\end{figure}

We show again a comparison of the actual temporal evolution of individual oscillators. Figure~\ref{f.Gaussian_TE1} shows results for the global synchronisation regime at $K=0.9$ and Figure~\ref{f.Gaussian_TE2} for the local synchronisation regime at $K=0.6$. In the case of the local synchronisation regime we assume that the oscillators which do not take part in the synchronised cluster are simply oscillating with their native frequencies and satisfy $\varphi_i(t) = (\alpha_0+t)\omega_i$. The temporal evolution is well described by the collective coordinate approach in both cases. It is clearly seen that, whereas the collective coordinate approach is able to capture the dynamics well of the well-entrained oscillators, it has difficulties describing the dynamics of the entrained extreme oscillators with large absolute native frequencies as seen in the insets of Figures~\ref{f.Gaussian_TE1} and \ref{f.Gaussian_TE2}. This discrepancy is due to the collective coordinate approach, as employed here, not taking into account the interaction with the drifting extreme oscillators.\\

%

%%%%%%%%%%%%%%%%%%%%%%%%%%%%%%%%%%%%%%%

\subsection{Bimodal distribution of native frequencies}
\label{sec-ATABimodal}
\begin{figure}
\begin{center}
\includegraphics[width=0.5\textwidth, height=0.2\textheight]{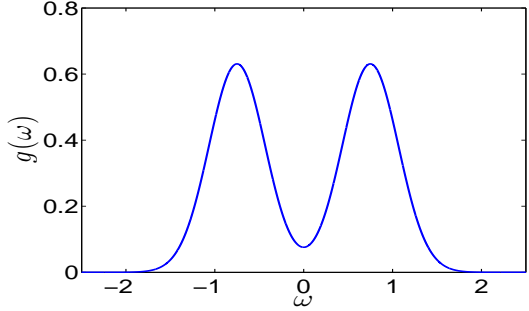}
\end{center}
\caption{Bimodal distribution $g(\omega)$ of native frequencies (\ref{e.bimodalpdf}) with $\sigma_\omega^2=0.1$ and $\Omega=0.75$.}
\label{f.bimodalpdf}
\end{figure}

In a third suite of experiments, we consider native frequencies which are distributed according to a bimodal distribution with maxima at $\omega=\pm \Omega$ and
\begin{align}
g(\omega)=\frac{1}{2 Z}\left(\exp\left(-\frac{(\omega+\Omega)^2}{2 \sigma_\omega^2}\right)+  \exp\left(-\frac{(\omega-\Omega)^2}{2 \sigma_\omega^2}\right)\right)\;.
\label{e.bimodalpdf}
\end{align}
We choose here $\sigma_\omega^2=0.1$ and $\Omega=0.75$. The bimodal distribution for these parameters is depicted in Figure~\ref{f.bimodalpdf}.
%The evolution equation (\ref{e.cc}) for the collective coordinate $\alpha(t)$ cannot be cast in a nice analytical form. 
%

\begin{figure}
\begin{center}
\includegraphics[width=0.5\textwidth, height=0.25\textheight]{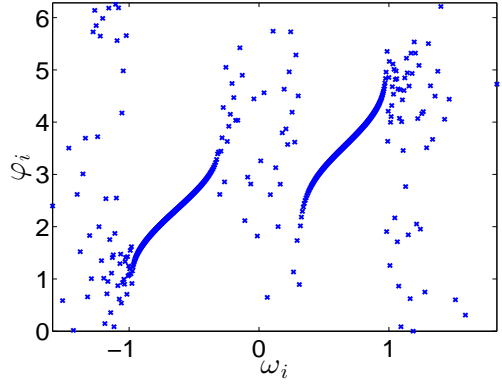}
\end{center}
\caption{Snapshot of the phases $\varphi_j$ for $K=1.2$ for the Kuramoto model with all-to-all coupling and bimodally distributed native frequencies with distribution (\ref{e.bimodalpdf}). One can see clearly the two partially synchronised clusters with frequencies centred around $\Omega=\pm 0.75$. The two clusters rotate with angular velocities of opposite sign forming a standing wave state.}
\label{f.twoclusters}
\end{figure}

\begin{figure}
\begin{center}
\includegraphics[width=0.5\textwidth, height=0.2\textheight]{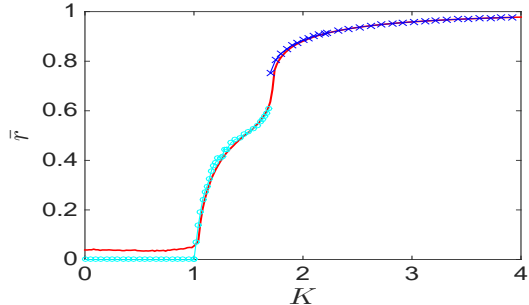}
\end{center}
\caption{Order parameter $\bar r$ as a function of the coupling strength $K$ for a network with bimodal native frequency distribution. Depicted are results from a direct numerical integration of the Kuramoto model (\ref{e.kuramoto}) with $N=500$ (continuous line, online red) and from the collective coordinate approach. We show results for global synchronisation (crosses, online blue) and for the standing wave state (open circles, online cyan).}
\label{f.bimodal_CollCoo}
\end{figure}

The synchronisation behaviour of Kuramoto networks with bimodal frequency distributions is more complex than in the two previous examples\cite{Kuramoto,BonillaEtAl92,Crawford94,MontbrioEtAl06,MartensEtAl09,PazoMontbrio09}. If the two peaks are sufficiently close together, the behaviour is, roughly speaking, as described in the unimodal case, discussed in the previous section, with local synchronisation being organised by oscillators with native frequencies closest to the mean frequency zero. However, when the peaks are sufficiently separated, a so called {\em{standing wave state \cite{Crawford94}}} occurs at some critical coupling strength $K=K_p$ whereby the oscillators with native frequencies close to the peak frequencies $\pm \Omega$ may synchronise and form two synchronised clusters which rotate with the same frequency but in the opposite direction. Upon increasing the coupling strength further, the oscillators will eventually globally synchronise at a critical coupling strength $K=K_c$ \cite{MartensEtAl09,PazoMontbrio09}. In Figure~\ref{f.twoclusters} we show a snapshot of the phases for the case $K_p<K<K_c$ where two partially synchronised clusters are established, centred around the nodes with $\omega_i=\pm\Omega$, respectively, which together form the standing wave state. In Figure~\ref{f.bimodal_CollCoo} we show the order parameter $\bar r$, where one can see clearly the standing wave state for $K_p<K<K_c$ and global synchronisation for $K>K_c$ with $K_p\approx 1.05$ and $K_c\approx1.7$.
 
First we apply our approach to the problem of global synchronisation, i.e. for $K>1.7$. In the thermodynamic limit the dynamic model for the collective coordinate (\ref{e.cc}) becomes
\begin{align}
\dot \alpha = 1 - & \frac{K}{\sigma_\omega^2 + \Omega^2} \exp(-\sigma_\omega^2\alpha^2)
\nonumber \\
& \times
\cos(\Omega \alpha)
\left( \sigma_\omega^2 \alpha  \cos(\Omega \alpha) + \Omega \sin(\Omega \alpha)\right) )\; .
\label{e.alphadotbimodal}
\end{align}
The equation for the order parameter (\ref{e.rhatgeneral}) can be evaluated in the thermodynamic limit to
\begin{align}
\hat r=\cos(\Omega \alpha)\exp\left(-\frac{\sigma_\omega^2 \alpha^2}{2}\right)\; .
\label{e.rhat_bimodal}
\end{align}
We have again omitted to write down the cumbersome expressions for the case of finite $N$, which nevertheless can readily be put into a numerical programme. 

\begin{figure}
\begin{center}
\includegraphics[width=0.5\textwidth, height=0.2\textheight]{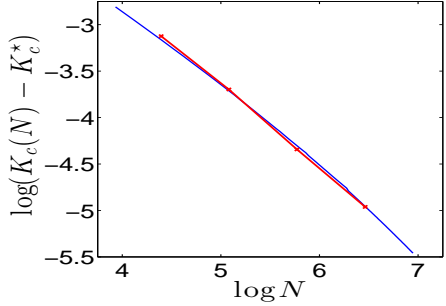}
\end{center}
\caption{Scaling of the critical coupling strength $K_c$ for the onset of global synchronisation as a function of the network size $N$ for a network with bimodal native frequency distribution. Depicted are results from a direct numerical integration of the Kuramoto model (\ref{e.kuramoto}) (crosses, online red) and from the collective coordinate approach (continuous line, online blue). The direct numerical simulations scale with a slope of $0.89$.}
\label{f.bimodal_finitesize}
\end{figure}

\begin{figure}
\begin{center}
\includegraphics[width=0.5\textwidth, height=0.2\textheight]{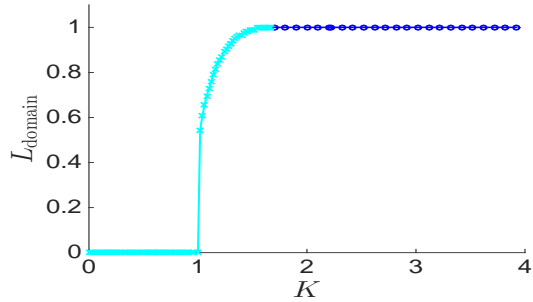}
\end{center}
\caption{Normalised length of phase synchronised domain as a function of the coupling strength for a network with bimodal native frequency distribution, calculated using the collective coordinate approach. The globally synchronised branch with $L_{\rm{domain}}=1$ tis preceded for $K<1.7$ by a standing wave state in which, for $K$ close to $1.7$, all oscillators are involved (i.e. $L_{\rm domain}=1$), but where the two partially synchronised clusters are not oscillating in phase.}
\label{f.bimodal_L}
\end{figure}

Figure~\ref{f.bimodal_CollCoo} shows a remarkable skill of the collective coordinate approach to describe the onset of global synchronisation and the order parameter $\bar r$. The critical coupling strength for the global synchronisation at $K_c=1.70$ is well captured. Furthermore, finite-size scaling can be described within our framework as shown in Figure~\ref{f.bimodal_finitesize} where we show a comparison of the critical coupling strength $K_c(N)$ as calculated via our collective coordinate approach for variable network sizes $N$ and results from direct simulations of the Kuramoto model (\ref{e.kuramoto}). As before we use as a proxy for the critical coupling strength the smallest value of the coupling strength $K$ such that $\bar r>0.8$. 
%The collective coordinate approach suggests that there is no clear power-law scaling.
\\ The normalised size $L_{\rm domain}$ of the globally synchronised cluster, which we determine as the largest number of nodes for which non-trivial stationary solutions $\alpha$ exist, is depicted in Figure~\ref{f.bimodal_L}. The smooth gradual decrease of $L_{\rm{domain}}$ with decreasing coupling strength $K$, is replaced here by a different behaviour caused by the standing wave state and the partial synchronisation of oscillators with native frequencies close to $\pm\Omega$. 

Oscillators with native frequencies $\omega \approx \pm \Omega$ near the maxima of the native frequency distribution experience local synchronisation similar to the case of unimodally distributed native frequencies discussed in Section~\ref{sec-ATAGaussian}. In the case of a bimodal frequency distribution this leads to two partially synchronised clusters - one with frequency close to $-\Omega$ and another one with frequency close to $+\Omega$ (cf. Figure~\ref{f.twoclusters}). With increasing coupling strength $K$ the two clusters grow in size and will start to interact before, upon further increasing $K$, they will merge at the onset of global synchronisation. We recall that this scenario only occurs provided the two peaks of the distribution of the native frequencies are sufficiently far separated allowing for a range in $K$ for which they can partially synchronise without interacting too strongly \cite{MartensEtAl09} to form the standing wave state. We now set out to describe the standing wave state in our collective coordinate approach.

In order to describe the effect of two partially synchronised clusters which rotate with non-uniform angular speeds of opposite direction we modify our ansatz and introduce a time-dependent phase function $f(t)$ as an additional collective coordinate. We split the phase oscillators into two groups, one group $\varphi_i^-$ describing the cluster centred around $-\Omega$, and one group $\varphi_i^+$ describing the cluster centred around $+\Omega$. We make the ansatz
\begin{align}
\varphi_i^{\pm} = \alpha(t)(\omega_i^{\pm}\mp \Omega ) \pm f(t)\; ,
\label{e.c3}
\end{align}
where $\omega_i^{\pm}$ are the native frequencies of the nodes participating in the cluster centred around $\pm\Omega$. Motivated by the results from direct simulations of the Kuramoto model we assume that each of the clusters consists of $N_2 \le N/2$ oscillators. Projecting the error onto the restricted subspace spanned by (\ref{e.c3}), i.e. onto $\partial \varphi_i^{\pm}/\partial \alpha = (\omega_i \mp \Omega)$ and onto $\partial \varphi_i^{\pm}/\partial f = \pm 1$, yields the desired evolution equations for $\alpha(t)$ and $f(t)$. Projecting onto $\partial \varphi_i^{-}/\partial \alpha$ and $\partial \varphi_i^{-}/\partial f$ yields
\begin{align}
\dot \alpha &= 1 + \frac{K}{2\Sigma^2}\frac{1}{N_2^2} \sum_i^{N_2}(\omega_i^-+\Omega)\sum_j^{N_2}\sin \alpha(\omega_j^--\omega_i^-) 
\nonumber\\
& - \frac{K}{2\Sigma^2}\frac{1}{N_2^2} \sum_i^{N_2}(\omega_i^-+\Omega)\sum_j^{N_2}\sin (\alpha(\omega_j^-+\omega_i^-)+2\alpha\Omega-2 f)
\label{e.ccN_bimodalN_alpha}\\
\dot f &= \Omega - \frac{1}{2}\frac{K}{N_2^2}\sum_{i,j}^{N_2}\sin \alpha(\omega_j^--\omega_i^-) 
\nonumber\\
&+ \frac{1}{2}\frac{K}{N_2^2} \sum_{i,j}^{N_2}\sin (\alpha(\omega_j^-+\omega_i^-)+2\alpha\Omega-2 f)
%\dot f &= -\Omega + \frac{1}{2}\frac{K}{\textstyle{\left(\frac{N}{2}\right)^2}}
\; ,
\label{e.ccN_bimodalN_f}
\end{align}
where here 
\begin{align}
\Sigma^2 = \frac{1}{N_2}\sum_j^{N_2}\left(\omega_j^-+\Omega\right)^2\; .
\label{e.sigma2_bimodal}
\end{align}
The sums are taken over indices representing the nodes within the clusters $\varphi_i^{-}$ (cf. (\ref{e.ccN2})). Due to symmetry projecting onto $\partial \varphi_i^{+}/\partial \alpha$ and $\partial \varphi_i^{+}/\partial f$ reduces to the same equation. The first sum in the right hand side of (\ref{e.ccN_bimodalN_alpha}) describes the interaction of oscillators within the partially synchronised cluster whereas the second sum describes the interaction of oscillators of one cluster with those of the respective other cluster.

In the thermodynamic limit $N\to \infty$, the evolution equations for the collective coordinates simplify in the case when $N_2=N/2$ to
\begin{align}
\dot \alpha &= 1 - \frac{K}{2} \exp{(-\alpha^2\sigma_\omega^2)}\alpha(1+\cos(2 f))
\label{e.alphadot_bimodal}
\\
\dot f &= \Omega - \frac{K}{2} \exp{(-\alpha^2\sigma_\omega^2)} \sin(2 f)
\; .
\label{e.ccN_bimodal}
\end{align}
Whereas in the case of global synchronisation the collective coordinate evolves to a stationary value, in the standing wave regime solutions of the system (\ref{e.ccN_bimodalN_alpha})-(\ref{e.ccN_bimodalN_f}) or (\ref{e.alphadot_bimodal})-(\ref{e.ccN_bimodal}) are oscillatory. These solutions can be found numerically. 
%The systems (\ref{e.ccN_bimodalN_alpha})-(\ref{e.ccN_bimodalN_f}) or (\ref{e.alphadot_bimodal})-(\ref{e.ccN_bimodal}) can be solved numerically to find stationary solutions for $\alpha$. 
The order parameter can then be calculated as an average of (\ref{e.rhatgeneral}) over one period $T_p$ of the phase function $f(t)$ and is given in the thermodynamic limit as
%For sufficiently separated peaks of the distribution $g(\omega)$ we can 
\begin{align}
\bar r = \frac{1}{T_p} \int_0^{T_p} \exp(-\frac{\alpha^2(t)\sigma_\omega^2}{2}) \cos f(t)\, dt\; ,
\label{e.rhat_bimodal_local}
\end{align}
In the thermodynamic limit the period $T_p$ can be determined analytically. Defining the collective coordinate $\bar \alpha$ as an average over the period $T_p$, the Adler equation (\ref{e.ccN_bimodal}) can be solved analytically as
\begin{align}
f(t) = \arctan \left( \frac{A+\sqrt{\Omega^2-A^2}\tan \sqrt{\Omega^2-A^2} t}{\Omega} \right)\; ,
\label{e.f0_bimodal}
\end{align}
with $A=(K/2) \exp(-\bar\alpha^2\sigma_\omega^2)$. The associated period $T_p$ is then defined as
%with $A={\textstyle{\frac{K}{2}}}\exp(-\alpha_s^2\sigma_\omega^2)$. The associated period $T_p$ is then defined as
\begin{align}
T_p=\int_0^{\pi}\frac{df}{\Omega-A\sin(2 f)} = \frac{\pi}{\sqrt{\Omega^2-A^2}} \; .
\label{e.Tp}
\end{align}
Note that because there are two counter-rotating clusters, the integration only goes to $\pi$ rather than to $2 \pi$.

In Figure~\ref{f.bimodal_CollCoo} we show results of the collective coordinate approach for the order parameter $\bar r$ as a function of the coupling strength $K$. In practice we first test for global synchronisation, and if this cannot be achieved for any domain length $L_{\rm domain}$, we test for the standing wave state. We have again allowed for local synchronisation whereby not all of the $N/2$ oscillators $\varphi_i^-$ are synchronised (cf. Figure~\ref{f.twoclusters}) analogously to (\ref{e.c2}) and (\ref{e.ccN2}). The onset of the standing wave state at $K_p=1.05$ is very well captured. The size of the synchronised clusters is shown in Figure~\ref{f.bimodal_L} where we count the total sum of locally synchronised oscillators $\varphi_i^-$ and $\varphi_i^+$ in the case of the standing wave state for $K<1.7$. 
%Finite-size scaling is recovered in Figure~\ref{f.bimodal_finitesize}. The collective coordinate approach suggests that there is no clear power-law scaling; the direct numerical simulations, however, show scaling with an exponent of $0.89$.
% is captured to a precision of less than $0.1\%$,  is approximated to a precision of less than $0.1\%$.

\begin{figure}
\begin{center}
\includegraphics[width=0.48\textwidth, height=0.25\textheight]{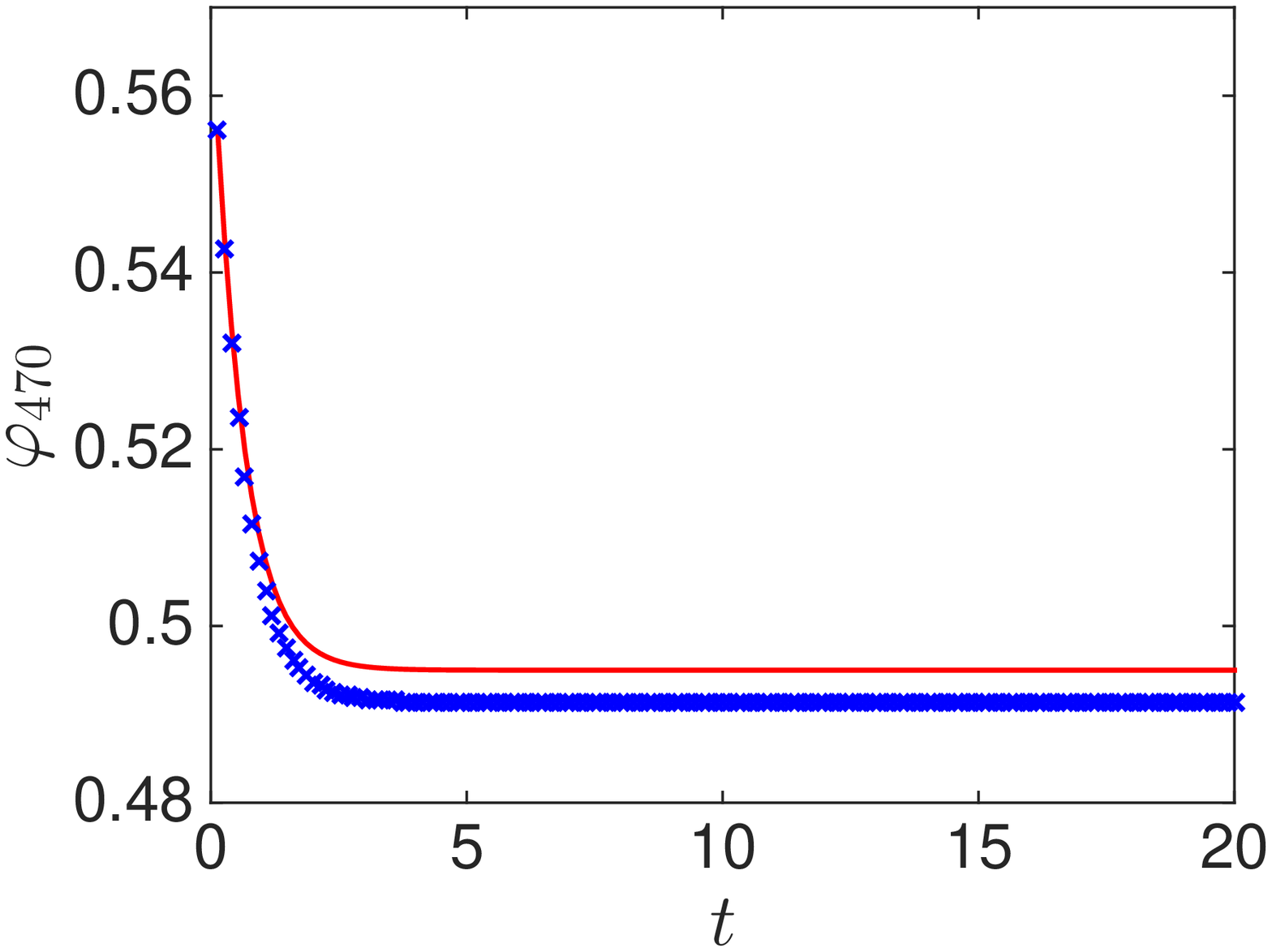}
\includegraphics[width=0.48\textwidth, height=0.25\textheight]{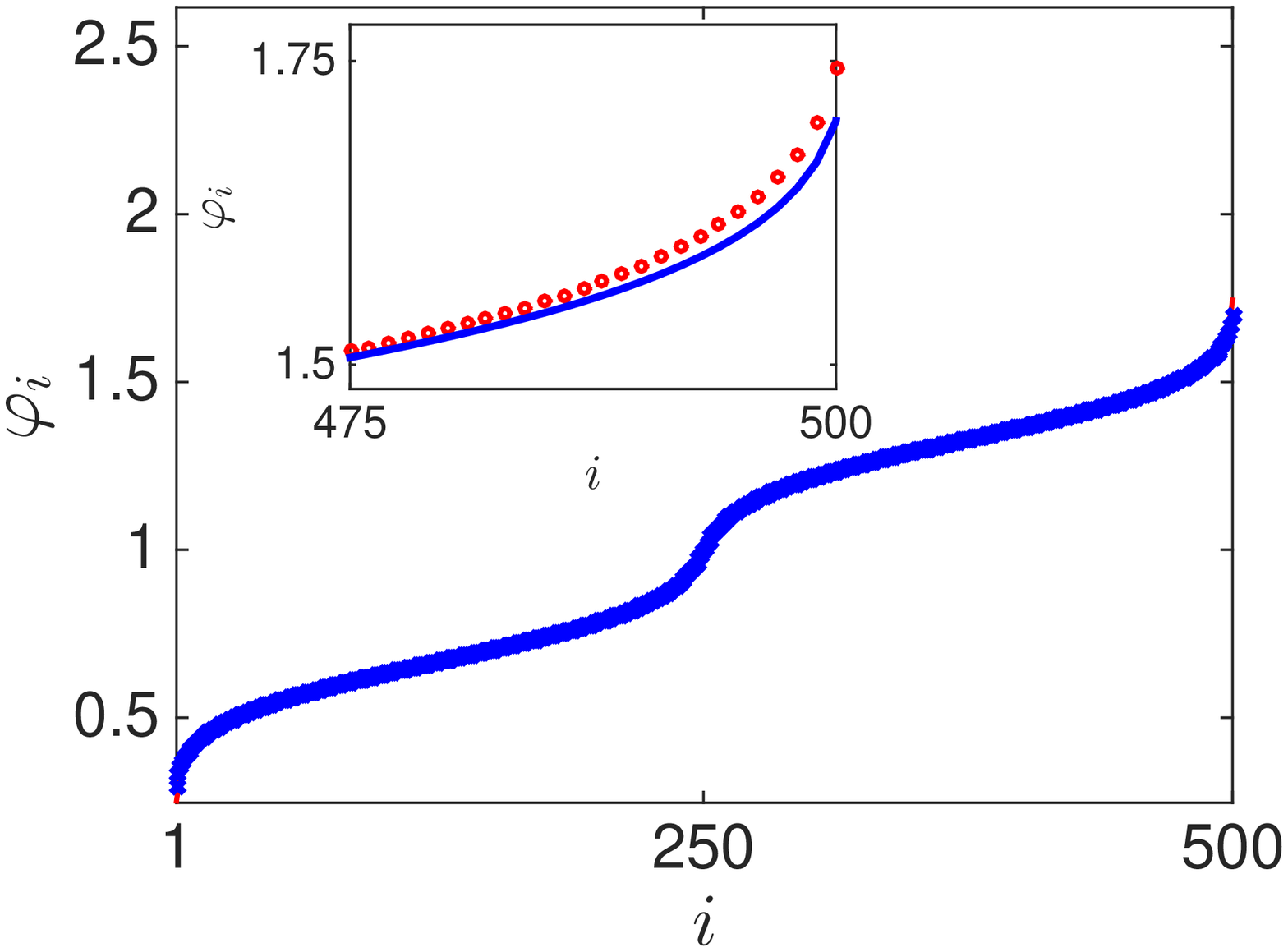}
\end{center}
\caption{Phases $\varphi(t)$ calculated from simulations of the full Kuramoto model (\ref{e.kuramoto}) (continuous lines or open circles, online red) and from the corresponding $1$-dimensional system (\ref{e.ccN}) for the collective coordinate with $\varphi_i = \alpha(t)\omega_i$ (crosses, online blue) for a network of $N=501$ oscillators with a bimodal distribution (\ref{e.bimodalpdf}) of the native frequencies at coupling strength $K=2.5$ corresponding to global synchronisation. Top: Temporal evolution of $\varphi_{470}(t)$ for initial conditions $\varphi_i(0) = \alpha_0 \omega_i$ with $\alpha_0 = 0.5$. The native frequency is $\omega_{470}=1.11$.Bottom: Snapshot of the phases $\varphi_i(T)$ at time $T=20$.}
%Open circles (online red) depict solution of the Kuramoto model (\ref{e.kuramoto}).
\label{f.Bimodal_TE1}
\end{figure}

\begin{figure}
\begin{center}
\includegraphics[width=0.48\textwidth, height=0.25\textheight]{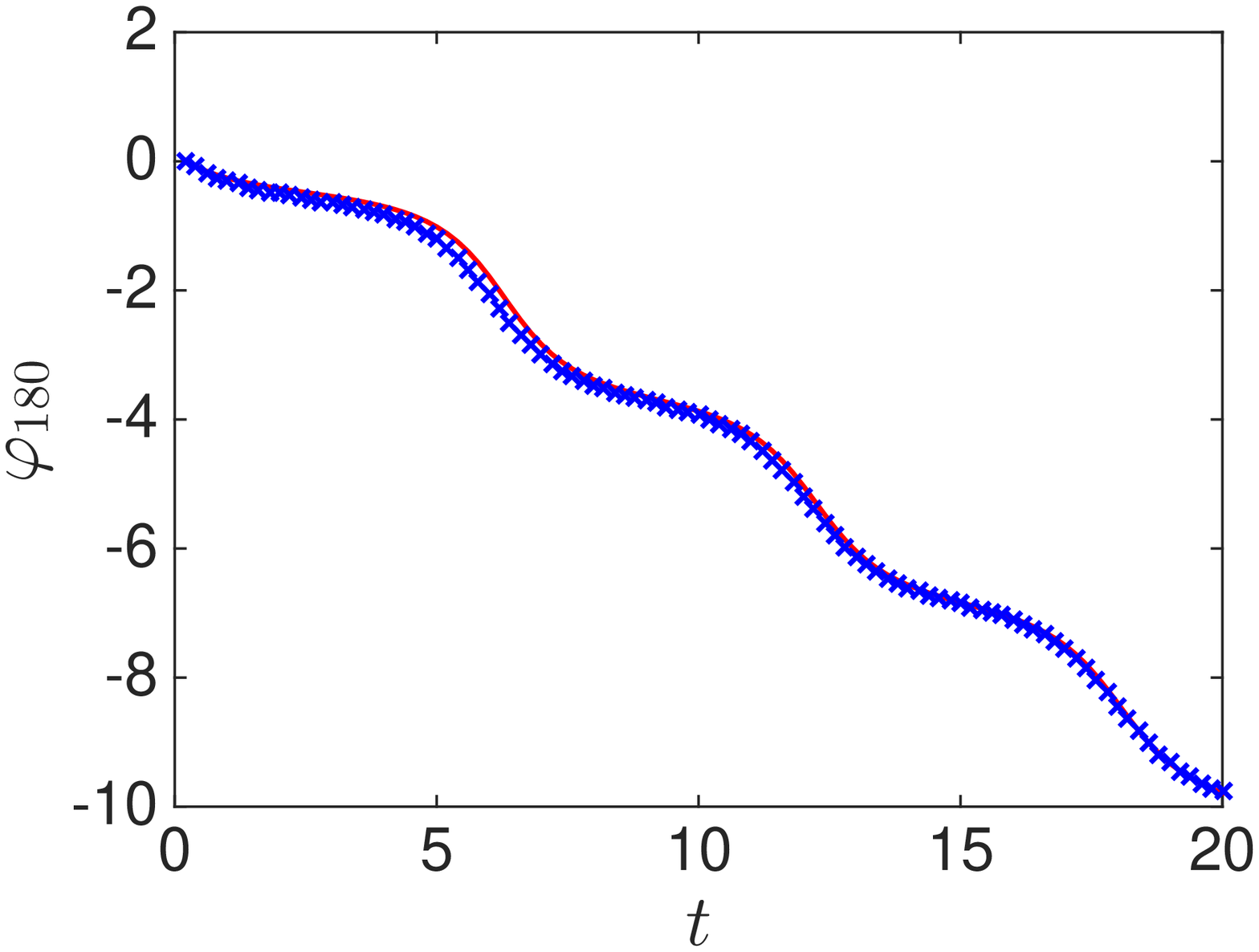}
\includegraphics[width=0.48\textwidth, height=0.25\textheight]{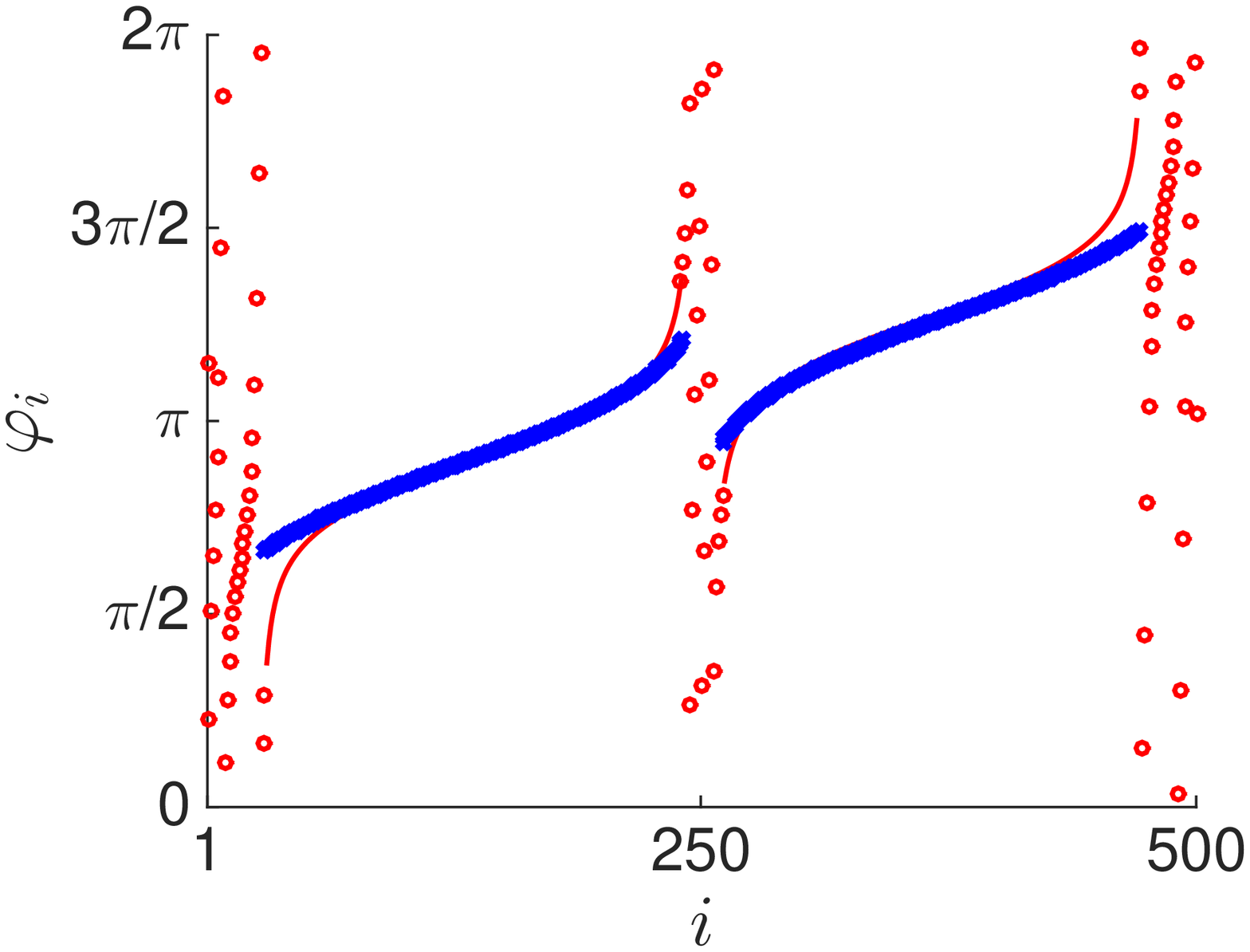}
\end{center}
\caption{Phases $\varphi(t)$ calculated from simulations of the full Kuramoto model (\ref{e.kuramoto}) (continuous lines or open circles, online red) and from the corresponding $2$-dimensional system (\ref{e.ccN_bimodal}) for the collective coordinate with $\varphi_i = \alpha(t)(\omega^{\pm}_i\mp\Omega) \pm f(t)$ (crosses, online blue) for a network of $N=501$ oscillators with a bimodal distribution (\ref{e.bimodalpdf}) of the native frequencies at coupling strength $K=1.3$ corresponding to the standing wave regime. The size of the two respective counter-rotating clusters is $N_2=210$. Top: Temporal evolution of $\varphi_{180}(t)$ for initial conditions $\varphi_i(0) = \alpha_0 (\omega_i + \Omega)$ with $\alpha_0 = 0.5$, i.e. $\alpha(0)=\alpha_0$ and $f(0)=0$. The native frequency is $\omega_{180}=-0.57$. Bottom: Snapshot of the phases $\varphi_i(T)$ at time $T=20$.}
% K=1.1 Nl2=190;
\label{f.Bimodal_TE2}
\end{figure}

In Figure~\ref{f.Bimodal_TE1} we show a comparison of the actual temporal evolution of individual oscillators in the global synchronisation regime at $K=2.5$ and in Figure~\ref{f.Bimodal_TE2} in the standing wave regime at $K=1.1$. The phases of the drifting oscillators which are not included within the collective coordinate analysis, are plotted simply by assuming that they are oscillating with their native frequencies. The actual phase dynamics of the synchronized oscillators is well described by our collective coordinate approach. One sees clearly the oscillatory behaviour of the phases in the standing wave regime which is caused by the interaction of the two counter-rotating clusters. The oscillation with period $T_p=5.8$ is well captured by the dynamics of the collective coordinates and matches approximately the analytically obtained period $T_p = 5.6$ if we use the sample mean and variance of the native frequencies instead of $\Omega$ and $\sigma_\omega^2$ in (\ref{e.Tp}).

%
% used qwe.m (and /or CollectiveCoordinates_AllToAll_Bimodal_twopopulations_parallel3.m) to produce:
%  CollCoord_AllToAll_Bimodal_twopop_Lo0p75_N501qwe.mat and/or
% CollCoord_AllToAll_Bimodal_twopop_Lo0p75_N501b_para3.mat

%
%

%%%%%%%%%%%%%%%%%%%%%%%%%%%%%%%%%%%%%%%
\section{Summary and Discussion}
\label{sec-summary}
The collective coordinate approach we propose allows for the description of networks of $N$ oscillators. The dimension $N$ is drastically reduced to a few $n$ judiciously chosen collective coordinates; here we presented examples with $n=1$ and $n=2$. The approach is not restricted to the thermodynamic limit of infinite network size and allows to study finite networks. The approach can be used to study the synchronisation behaviour of networks, both global and partial, and determine the order parameter and the size of the synchronised clusters. Besides capturing this collective behaviour of oscillators the collective coordinate approach also is able to resolve the temporal evolution of individual oscillators for a wide range of coupling strength.\\

We have corroborated our approach for the Kuramoto model with all-to-all coupling in numerical simulations for different distributions of the native frequencies. We found good agreement of our reduced $1$-dimensional model (or $2$-dimensional model in the case of bimodal native frequency distributions) with the full $N$-dimensional system. In particular, the behaviour of the order parameter was well captured and the approach was able to describe soft second-order as well as explosive first-order transitions to synchronisation. We have illustrated that the collective coordinate approach reproduces finite size scalings of the full system. Furthermore, the approach allowed to describe the interplay between a standing wave state involving partially synchronised counter-rotating clusters and global synchronisation in networks with bimodal distribution of native frequencies. We have shown that the collective coordinates are able to capture the dynamics of individual oscillators which is a much stronger form of approximation than just reproducing the collective behaviour.\\

It is pertinent to caution that the method is by no means rigorous. The choice of collective coordinates is so far limited to {\em a priori} information obtained from direct numerical simulations of the full dynamical network. We have seen that transitory temporal evolution of oscillators in a Kuramoto model is only well described by the collective coordinate method provided the initial conditions are sufficiently close to the synchronisation manifold. Furthermore, the temporal evolution of individual oscillators at the edge of a synchronised cluster is not accurately captured. To put our ansatz on a firm theoretical footing which allows to describe its limitations is an open question.\\ 

From a practical point of view, there are several issues which require further attention and which we plan to pursue in future research. First of all, whereas the general framework of collective coordinates is formulated for general network topologies, we have only presented numerical results for the case of an all-to-all coupling. It is an interesting and important question to see whether the success of the method translates to more complex network topologies.\\ Second, it is by no means clear that our ansatz captures all possible attractors of the full dynamical system. For example, there are examples of networks where the Ott-Antonson method of reduction \cite{OttAntonson08} does not account for the actual dynamical behaviour observed in these networks (see the discussion in \citet{MartensEtAl09}). In particular, chaotic dynamics is excluded from their framework. The collective coordinate approach is, in principle, capable of recovering chaotic dynamics by considering at least three collective coordinates. To test whether it actually is able to describe more complex dynamic behaviour is an interesting avenue to pursue.\\ Thirdly, the success in describing the interaction between two partially synchronised clusters in the case of bimodally distributed native frequencies suggests that collective coordinates may be used to reduce complex networks involving several clusters or communities.\\ Fourthly, as we have seen in the numerical simulation, the collective coordinate approach does not capture the interaction between the drifter oscillators and the synchronised oscillators. This leads to the collective coordinate behaviour not being able to accurately capture the oscillators which sit on the edge of the cluster. At a next step one can extend the approach to include drifters.\\

%%%%%%%%%%%%%%%%%%%%%%%%%%%%%%%%%%%%%%%
\vspace{0.5cm}
{\bf{Acknowledgments:} }
I thank Hyunggyu Park for valuable discussions. I acknowledge support from the Australian Research Council.

%%%%%%%%%%%%%%%%%%%%%%%%%%%%%%%%%%%%%%%%%%%%

%\bibliographystyle{natbib}
%\bibliography{bibliography}

\end{document}